%%%%%%%%%%%%%%%%%%%%%%%%%%%%%%%%%%%%%%%%%%%%%%%%%%%%%%%%%%%%%%%%%%%%%
%
%
%                                      Andrea PASQUINUCCI, 1988
%              PANDA.TEX               S.I.S.S.A., Trieste, Italy
%                                      (Revised 1991, Princeton, USA)
%
%--------------------------------------------------------------------
%
%    These are TEX macros. They work with PLAIN TEX (the basis
%    version of TEX). The only problem can be with the double-page
%    format since it depends on the type of software and laserwriter
%    you use to print, so I cannot guarantee that the double-page
%    format will work properly. Double-page MUST be printed in
%    LANDSCAPE orientation. (You shouldn't have troubles with fonts;
%    if you do, please let me know.)
%
%--------------------------------------------------------------------
%
%                     INTERACTIVE SECTION
%
%--------------------------------------------------------------------
%
\def\standardrisposta{s }\def\reducedrisposta{r }
\def\mplarisposta{mpla }\def\zerorisposta{z }
\def\doublerisposta{d }\def\cartarisposta{e }\def\amsrisposta{y }
\newcount\ingrandimento \newcount\sinnota \newcount\dimnota
\newcount\unoduecol \newdimen\collhsize \newdimen\tothsize
\newdimen\fullhsize \newcount\controllorisposta \sinnota=1
\newskip\infralinea  \global\controllorisposta=0
\immediate\write16 { ********  Welcome to PANDA macros (Plain TeX,
AP, 1991) ******** }
%\immediate\write16 { You'll have to answer a few questions in
%lowercase.}
%\message{>  Do you want it in double-page (d), reduced (r)
%or standard format (s) ? }\read-1 to\risposta
%
%\message{>  Do you want it in USA A4 (u) or EUROPEAN A4
%(e) paper size ? }\read-1 to\srisposta
%
%\message{>  Do you have AMSFonts 2.0 (math) fonts (y/n) ? }
%\read-1 to\arisposta
%
%--------------------------------------------------------------------
%
%             END INTERACTIVE SECTION - PAGE FORMATTING
%
%--------------------------------------------------------------------
%       The following parameters define defaults to the interactive
%       session.  At the moment I have set EUROPEAN and MATH FONTS
%
\def\risposta{s } 
\def\srisposta{e } 
\def\arisposta{y }
\ifx\risposta\standardrisposta \ingrandimento=1200
\message {>> This will come out UNREDUCED << }
\dimnota=2 \unoduecol=1 \global\controllorisposta=1 \fi
\ifx\risposta\reducedrisposta \ingrandimento=1095 \dimnota=1
\unoduecol=1  \global\controllorisposta=1
\message {>> This will come out REDUCED << } \fi
\ifx\risposta\doublerisposta \ingrandimento=1000 \dimnota=2
\unoduecol=2

\message {>> You must print this in
LANDSCAPE orientation << } \global\controllorisposta=1 \fi
\ifx\risposta\mplarisposta \ingrandimento=1000 \dimnota=1
\message {>> Mod. Phys. Lett. A format << }
\unoduecol=1 \global\controllorisposta=1 \fi
\ifx\risposta\zerorisposta \ingrandimento=1000 \dimnota=2
\message {>> Zero Magnification format << }
\unoduecol=1 \global\controllorisposta=1 \fi
\ifnum\controllorisposta=0  \ingrandimento=1200
\message {>>> ERROR IN INPUT, I ASSUME STANDARD
UNREDUCED FORMAT <<< }  \dimnota=2 \unoduecol=1 \fi
\magnification=\ingrandimento
%
%--------------------------------------------------------------------
%
%                        PARAMETERS SETTING
%
%  You can modify these parameters at your will (and resposability)
%--------------------------------------------------------------------
%
\newdimen\eucolumnsize \newdimen\eudoublehsize \newdimen\eudoublevsize
\newdimen\uscolumnsize \newdimen\usdoublehsize \newdimen\usdoublevsize
\newdimen\eusinglehsize \newdimen\eusinglevsize \newdimen\ussinglehsize
\newskip\standardbaselineskip \newdimen\ussinglevsize
\newskip\reducedbaselineskip \newskip\doublebaselineskip
\eucolumnsize=12.0truecm    % column h-size for european doublepage
                            % (12.0treucm default)
\eudoublehsize=25.5truecm   % sheet h-size for european duoblepage
                            % (25.5treucm default)
\eudoublevsize=6.7truein    % sheet v-size for european doublepage
                            % (6.5treuin default  or 17truecm?)
\uscolumnsize=4.4truein     % column h-size for american doublepage
                            % (4.4treuin default)
\usdoublehsize=9.4truein    % sheet h-size for american duoblepage
                            % (9.4treuin default)
\usdoublevsize=6.8truein    % sheet v-size for american doublepage
                            % (6.8treuin default)
\eusinglehsize=6.5truein    % sheet h-size for european singlepage
                            % (6.5truein default)
\eusinglevsize=24truecm     % sheet v-size for european singlepage
                            % (24truecm default)
\ussinglehsize=6.5truein    % sheet h-size for american singlepage
                            % (6.5truein default)
\ussinglevsize=8.9truein    % sheet v-size for american singlepage
                            % (8.9truein default)
\standardbaselineskip=16pt plus.2pt  % baselineskip for standard
                                     % format (16pt default)
\reducedbaselineskip=14pt plus.2pt   % baselineskip for reduced
                                     % format (14pt default)
\doublebaselineskip=12pt plus.2pt    % baselineskip for doublepage
                                     % format (12pt default)
%
%  \Portoffset and \Landoffset define the horizontal and vertical
%  offsets respectively for portrait and landscape modes. Example:
%  \def\Portoffset{\voffset=.4truein\hoffset=.125truein}
%
\def\Portoffset{}
\def\Landoffset{\voffset=-.2truein}
\ifx\risposta\mplarisposta \def\Portoffset{\hoffset=1.8truecm} \fi
%
%  \Landspec defines the \special command that sets the printer
%  to landscape mode without need to specify it directly in the
%  TeX to postscript translator (the command is site dependent).
%  Example: \def\Landspec{\special{ps: landscape}}
%
\def\Landspec{}
\tolerance=10000
\parskip=0pt plus2pt  \leftskip=0pt \rightskip=0pt
%
%   Do not modify anything of what follows
%                       (unless you know what you are doing!)
%----------------------------------------------------------------------
%
\ifx\risposta\standardrisposta \infralinea=\standardbaselineskip \fi
\ifx\risposta\reducedrisposta  \infralinea=\reducedbaselineskip \fi
\ifx\risposta\doublerisposta   \infralinea=\doublebaselineskip \fi
\ifx\risposta\mplarisposta     \infralinea=13pt \fi
\ifx\risposta\zerorisposta     \infralinea=12pt plus.2pt\fi
\ifnum\controllorisposta=0    \infralinea=\standardbaselineskip \fi
\ifx\risposta\doublerisposta   \Landoffset \else \Portoffset \fi
\ifx\risposta\doublerisposta \ifx\srisposta\cartarisposta
\tothsize=\eudoublehsize \collhsize=\eucolumnsize
\vsize=\eudoublevsize  \else  \tothsize=\usdoublehsize
\collhsize=\uscolumnsize \vsize=\usdoublevsize \fi \else
\ifx\srisposta\cartarisposta \tothsize=\eusinglehsize
\vsize=\eusinglevsize \else  \tothsize=\ussinglehsize
\vsize=\ussinglevsize \fi \collhsize=4.4truein \fi
\ifx\risposta\mplarisposta \tothsize=5.0truein
\vsize=7.8truein \collhsize=4.4truein \fi
%
%--------------------------------------------------------------------
%
%                            FONTS
%
%--------------------------------------------------------------------
%
\newcount\contaeuler \newcount\contacyrill \newcount\contaams
\font\ninerm=cmr9  \font\eightrm=cmr8  \font\sixrm=cmr6
\font\ninei=cmmi9  \font\eighti=cmmi8  \font\sixi=cmmi6
\font\ninesy=cmsy9  \font\eightsy=cmsy8  \font\sixsy=cmsy6
\font\ninebf=cmbx9  \font\eightbf=cmbx8  \font\sixbf=cmbx6
\font\ninett=cmtt9  \font\eighttt=cmtt8  \font\nineit=cmti9
\font\eightit=cmti8 \font\ninesl=cmsl9  \font\eightsl=cmsl8
\skewchar\ninei='177 \skewchar\eighti='177 \skewchar\sixi='177
\skewchar\ninesy='60 \skewchar\eightsy='60 \skewchar\sixsy='60
\hyphenchar\ninett=-1 \hyphenchar\eighttt=-1 \hyphenchar\tentt=-1
\def\bfmath{\cmmib}                 % math italic bold \bfmath
\font\tencmmib=cmmib10  \newfam\cmmibfam  \skewchar\tencmmib='177
                  % math bold (cal) symbols
\font\tencmbsy=cmbsy10  \newfam\cmbsyfam  \skewchar\tencmbsy='60
\def\scaps{\cmcsc}                 % small caps (uppercase)
\font\tencmcsc=cmcsc10  \newfam\cmcscfam
\ifnum\ingrandimento=1095

\font\capsone=cmcsc10 at 10.95pt \font\capstwo=cmcsc10 at 13.145pt

\else

\font\capsone=cmcsc10 at 12pt \font\capstwo=cmcsc10 at 14.4pt
\fi

\def\ttaarr{\bf}		% chapter titles' font
\def\ppaarr{\sl}		% section titles' font

%
     % inch-high caps (enormous)
%
%   AMS fonts (this works only if you have at least the 2.0
%              version of AMSFonts, otherwise say no)
%
\newfam\eufmfam \newfam\msamfam \newfam\msbmfam \newfam\eufbfam
\def\Loadeulerfonts{\global\contaeuler=1 \ifx\arisposta\amsrisposta
\font\teneufm=eufm10              %  \eufm   Gothic (or Euler)
\font\eighteufm=eufm8 \font\nineeufm=eufm9 \font\sixeufm=eufm6
\font\seveneufm=eufm7  \font\fiveeufm=eufm5
\font\teneufb=eufb10              %  \eufb   Bold Gothic (or Euler)
\font\eighteufb=eufb8 \font\nineeufb=eufb9 \font\sixeufb=eufb6
\font\seveneufb=eufb7  \font\fiveeufb=eufb5
\font\teneurm=eurm10              %  \eurm   Roman Gothic (or Euler)
\font\eighteurm=eurm8 \font\nineeurm=eurm9
\font\teneurb=eurb10              %  \eurb   Roman Bold Gothic
\font\eighteurb=eurb8 \font\nineeurb=eurb9
\font\teneusm=eusm10              %  \eusm   Slanted Capital Gothic
\font\eighteusm=eusm8 \font\nineeusm=eusm9
\font\teneusb=eusb10              %\eusb Slanted Capital Bold Gothic
\font\eighteusb=eusb8 \font\nineeusb=eusb9
\else \def\eufm{\tt} \def\eufb{\tt} \def\eurm{\tt} \def\eurb{\tt}
\def\eusm{\tt} \def\eusb{\tt}    \fi}
\def\loadeuler{\Loadeulerfonts\tenpoint}
\def\loadamsmath{\global\contaams=1 \ifx\arisposta\amsrisposta
\font\tenmsam=msam10 \font\ninemsam=msam9 \font\eightmsam=msam8
\font\sevenmsam=msam7 \font\sixmsam=msam6 \font\fivemsam=msam5
\font\tenmsbm=msbm10 \font\ninemsbm=msbm9 \font\eightmsbm=msbm8
\font\sevenmsbm=msbm7 \font\sixmsbm=msbm6 \font\fivemsbm=msbm5
\else \def\msbm{\bf} \fi \def\Bbb{\msbm} \def\symbl{\msam} \tenpoint}
\def\loadcyrill{\global\contacyrill=1 \ifx\arisposta\amsrisposta
\font\tenwncyr=wncyr10 \font\ninewncyr=wncyr9 \font\eightwncyr=wncyr8
\font\tenwncyb=wncyr10 \font\ninewncyb=wncyr9 \font\eightwncyb=wncyr8
\font\tenwncyi=wncyr10 \font\ninewncyi=wncyr9 \font\eightwncyi=wncyr8
\else \def\cyrill{\sl} \def\cyrilb{\sl} \def\cyrili{\sl} \fi\tenpoint}
\ifx\arisposta\amsrisposta
\font\sevenex=cmex7               %  reduced math symbols
\font\eightex=cmex8  \font\nineex=cmex9
\font\ninecmmib=cmmib9   \font\eightcmmib=cmmib8
\font\sevencmmib=cmmib7 \font\sixcmmib=cmmib6
\font\fivecmmib=cmmib5   \skewchar\ninecmmib='177
\skewchar\eightcmmib='177  \skewchar\sevencmmib='177
\skewchar\sixcmmib='177   \skewchar\fivecmmib='177
\font\ninecmbsy=cmbsy9    \font\eightcmbsy=cmbsy8
\font\sevencmbsy=cmbsy7  \font\sixcmbsy=cmbsy6
\font\fivecmbsy=cmbsy5   \skewchar\ninecmbsy='60
\skewchar\eightcmbsy='60  \skewchar\sevencmbsy='60
\skewchar\sixcmbsy='60    \skewchar\fivecmbsy='60
\font\ninecmcsc=cmcsc9    \font\eightcmcsc=cmcsc8     \else
\def\cmmib{\fam\cmmibfam\tencmmib}\textfont\cmmibfam=\tencmmib
\scriptfont\cmmibfam=\tencmmib \scriptscriptfont\cmmibfam=\tencmmib
\def\cmbsy{\fam\cmbsyfam\tencmbsy} \textfont\cmbsyfam=\tencmbsy
\scriptfont\cmbsyfam=\tencmbsy \scriptscriptfont\cmbsyfam=\tencmbsy
\scriptfont\cmcscfam=\tencmcsc \scriptscriptfont\cmcscfam=\tencmcsc
\def\cmcsc{\fam\cmcscfam\tencmcsc} \textfont\cmcscfam=\tencmcsc \fi
\catcode`@=11
\newskip\ttglue
\gdef\tenpoint{\def\rm{\fam0\tenrm}
  \textfont0=\tenrm \scriptfont0=\sevenrm \scriptscriptfont0=\fiverm
  \textfont1=\teni \scriptfont1=\seveni \scriptscriptfont1=\fivei
  \textfont2=\tensy \scriptfont2=\sevensy \scriptscriptfont2=\fivesy
  \textfont3=\tenex \scriptfont3=\tenex \scriptscriptfont3=\tenex
  \def\mcal{\fam2 \tensy}  \def\mmit{\fam1 \teni}
  \textfont\itfam=\tenit \def\it{\fam\itfam\tenit}
  \textfont\slfam=\tensl \def\sl{\fam\slfam\tensl}
  \textfont\ttfam=\tentt \scriptfont\ttfam=\eighttt
  \scriptscriptfont\ttfam=\eighttt  \def\tt{\fam\ttfam\tentt}
  \textfont\bffam=\tenbf \scriptfont\bffam=\sevenbf
  \scriptscriptfont\bffam=\fivebf \def\bf{\fam\bffam\tenbf}
     \ifx\arisposta\amsrisposta    \ifnum\contaeuler=1
  \textfont\eufmfam=\teneufm \scriptfont\eufmfam=\seveneufm
  \scriptscriptfont\eufmfam=\fiveeufm \def\eufm{\fam\eufmfam\teneufm}
  \textfont\eufbfam=\teneufb \scriptfont\eufbfam=\seveneufb
  \scriptscriptfont\eufbfam=\fiveeufb \def\eufb{\fam\eufbfam\teneufb}
  \def\eurm{\teneurm} \def\eurb{\teneurb} \def\eusm{\teneusm}
  \def\eusb{\teneusb}    \fi    \ifnum\contaams=1
  \textfont\msamfam=\tenmsam \scriptfont\msamfam=\sevenmsam
  \scriptscriptfont\msamfam=\fivemsam \def\msam{\fam\msamfam\tenmsam}
  \textfont\msbmfam=\tenmsbm \scriptfont\msbmfam=\sevenmsbm
  \scriptscriptfont\msbmfam=\fivemsbm \def\msbm{\fam\msbmfam\tenmsbm}
     \fi      \ifnum\contacyrill=1     \def\cyrill{\tenwncyr}
  \def\cyrilb{\tenwncyb}  \def\cyrili{\tenwncyi}         \fi
  \textfont3=\tenex \scriptfont3=\sevenex \scriptscriptfont3=\sevenex
  \def\cmmib{\fam\cmmibfam\tencmmib} \scriptfont\cmmibfam=\sevencmmib
  \textfont\cmmibfam=\tencmmib  \scriptscriptfont\cmmibfam=\fivecmmib
  \def\cmbsy{\fam\cmbsyfam\tencmbsy} \scriptfont\cmbsyfam=\sevencmbsy
  \textfont\cmbsyfam=\tencmbsy  \scriptscriptfont\cmbsyfam=\fivecmbsy
  \def\cmcsc{\fam\cmcscfam\tencmcsc} \scriptfont\cmcscfam=\eightcmcsc
  \textfont\cmcscfam=\tencmcsc \scriptscriptfont\cmcscfam=\eightcmcsc
     \fi            \tt \ttglue=.5em plus.25em minus.15em
  \normalbaselineskip=12pt
  \setbox\strutbox=\hbox{\vrule height8.5pt depth3.5pt width0pt}
  \let\sc=\eightrm \let\big=\tenbig   \normalbaselines
  \baselineskip=\infralinea  \rm}
\gdef\ninepoint{\def\rm{\fam0\ninerm}
  \textfont0=\ninerm \scriptfont0=\sixrm \scriptscriptfont0=\fiverm
  \textfont1=\ninei \scriptfont1=\sixi \scriptscriptfont1=\fivei
  \textfont2=\ninesy \scriptfont2=\sixsy \scriptscriptfont2=\fivesy
  \textfont3=\tenex \scriptfont3=\tenex \scriptscriptfont3=\tenex
  \def\mcal{\fam2 \ninesy}  \def\mmit{\fam1 \ninei}
  \textfont\itfam=\nineit \def\it{\fam\itfam\nineit}
  \textfont\slfam=\ninesl \def\sl{\fam\slfam\ninesl}
  \textfont\ttfam=\ninett \scriptfont\ttfam=\eighttt
  \scriptscriptfont\ttfam=\eighttt \def\tt{\fam\ttfam\ninett}
  \textfont\bffam=\ninebf \scriptfont\bffam=\sixbf
  \scriptscriptfont\bffam=\fivebf \def\bf{\fam\bffam\ninebf}
     \ifx\arisposta\amsrisposta  \ifnum\contaeuler=1
  \textfont\eufmfam=\nineeufm \scriptfont\eufmfam=\sixeufm
  \scriptscriptfont\eufmfam=\fiveeufm \def\eufm{\fam\eufmfam\nineeufm}
  \textfont\eufbfam=\nineeufb \scriptfont\eufbfam=\sixeufb
  \scriptscriptfont\eufbfam=\fiveeufb \def\eufb{\fam\eufbfam\nineeufb}
  \def\eurm{\nineeurm} \def\eurb{\nineeurb} \def\eusm{\nineeusm}
  \def\eusb{\nineeusb}     \fi   \ifnum\contaams=1
  \textfont\msamfam=\ninemsam \scriptfont\msamfam=\sixmsam
  \scriptscriptfont\msamfam=\fivemsam \def\msam{\fam\msamfam\ninemsam}
  \textfont\msbmfam=\ninemsbm \scriptfont\msbmfam=\sixmsbm
  \scriptscriptfont\msbmfam=\fivemsbm \def\msbm{\fam\msbmfam\ninemsbm}
     \fi       \ifnum\contacyrill=1     \def\cyrill{\ninewncyr}
  \def\cyrilb{\ninewncyb}  \def\cyrili{\ninewncyi}         \fi
  \textfont3=\nineex \scriptfont3=\sevenex \scriptscriptfont3=\sevenex
  \def\cmmib{\fam\cmmibfam\ninecmmib}  \textfont\cmmibfam=\ninecmmib
  \scriptfont\cmmibfam=\sixcmmib \scriptscriptfont\cmmibfam=\fivecmmib
  \def\cmbsy{\fam\cmbsyfam\ninecmbsy}  \textfont\cmbsyfam=\ninecmbsy
  \scriptfont\cmbsyfam=\sixcmbsy \scriptscriptfont\cmbsyfam=\fivecmbsy
  \def\cmcsc{\fam\cmcscfam\ninecmcsc} \scriptfont\cmcscfam=\eightcmcsc
  \textfont\cmcscfam=\ninecmcsc \scriptscriptfont\cmcscfam=\eightcmcsc
     \fi            \tt \ttglue=.5em plus.25em minus.15em
  \normalbaselineskip=11pt
  \setbox\strutbox=\hbox{\vrule height8pt depth3pt width0pt}
  \let\sc=\sevenrm \let\big=\ninebig \normalbaselines\rm}
\gdef\eightpoint{\def\rm{\fam0\eightrm}
  \textfont0=\eightrm \scriptfont0=\sixrm \scriptscriptfont0=\fiverm
  \textfont1=\eighti \scriptfont1=\sixi \scriptscriptfont1=\fivei
  \textfont2=\eightsy \scriptfont2=\sixsy \scriptscriptfont2=\fivesy
  \textfont3=\tenex \scriptfont3=\tenex \scriptscriptfont3=\tenex
  \def\mcal{\fam2 \eightsy}  \def\mmit{\fam1 \eighti}
  \textfont\itfam=\eightit \def\it{\fam\itfam\eightit}
  \textfont\slfam=\eightsl \def\sl{\fam\slfam\eightsl}
  \textfont\ttfam=\eighttt \scriptfont\ttfam=\eighttt
  \scriptscriptfont\ttfam=\eighttt \def\tt{\fam\ttfam\eighttt}
  \textfont\bffam=\eightbf \scriptfont\bffam=\sixbf
  \scriptscriptfont\bffam=\fivebf \def\bf{\fam\bffam\eightbf}
     \ifx\arisposta\amsrisposta   \ifnum\contaeuler=1
  \textfont\eufmfam=\eighteufm \scriptfont\eufmfam=\sixeufm
  \scriptscriptfont\eufmfam=\fiveeufm \def\eufm{\fam\eufmfam\eighteufm}
  \textfont\eufbfam=\eighteufb \scriptfont\eufbfam=\sixeufb
  \scriptscriptfont\eufbfam=\fiveeufb \def\eufb{\fam\eufbfam\eighteufb}
  \def\eurm{\eighteurm} \def\eurb{\eighteurb} \def\eusm{\eighteusm}
  \def\eusb{\eighteusb}       \fi    \ifnum\contaams=1
  \textfont\msamfam=\eightmsam \scriptfont\msamfam=\sixmsam
  \scriptscriptfont\msamfam=\fivemsam \def\msam{\fam\msamfam\eightmsam}
  \textfont\msbmfam=\eightmsbm \scriptfont\msbmfam=\sixmsbm
  \scriptscriptfont\msbmfam=\fivemsbm \def\msbm{\fam\msbmfam\eightmsbm}
     \fi       \ifnum\contacyrill=1     \def\cyrill{\eightwncyr}
  \def\cyrilb{\eightwncyb}  \def\cyrili{\eightwncyi}         \fi
  \textfont3=\eightex \scriptfont3=\sevenex \scriptscriptfont3=\sevenex
  \def\cmmib{\fam\cmmibfam\eightcmmib}  \textfont\cmmibfam=\eightcmmib
  \scriptfont\cmmibfam=\sixcmmib \scriptscriptfont\cmmibfam=\fivecmmib
  \def\cmbsy{\fam\cmbsyfam\eightcmbsy}  \textfont\cmbsyfam=\eightcmbsy
  \scriptfont\cmbsyfam=\sixcmbsy \scriptscriptfont\cmbsyfam=\fivecmbsy
  \def\cmcsc{\fam\cmcscfam\eightcmcsc} \scriptfont\cmcscfam=\eightcmcsc
  \textfont\cmcscfam=\eightcmcsc \scriptscriptfont\cmcscfam=\eightcmcsc
     \fi             \tt \ttglue=.5em plus.25em minus.15em
  \normalbaselineskip=9pt
  \setbox\strutbox=\hbox{\vrule height7pt depth2pt width0pt}
  \let\sc=\sixrm \let\big=\eightbig \normalbaselines\rm }
\gdef\tenbig#1{{\hbox{$\left#1\vbox to8.5pt{}\right.\n@space$}}}
\gdef\ninebig#1{{\hbox{$\textfont0=\tenrm\textfont2=\tensy
   \left#1\vbox to7.25pt{}\right.\n@space$}}}
\gdef\eightbig#1{{\hbox{$\textfont0=\ninerm\textfont2=\ninesy
   \left#1\vbox to6.5pt{}\right.\n@space$}}}
 %for 10-pt math in 9-pt territory
\def\alternativefont#1#2{\ifx\arisposta\amsrisposta \relax \else
\xdef#1{#2} \fi}
\global\contaeuler=0 \global\contacyrill=0 \global\contaams=0
%
%--------------------------------------------------------------------
%
%                            MACROS
%
%--------------------------------------------------------------------
%
\newbox\fotlinebb \newbox\hedlinebb \newbox\leftcolumn
\gdef\makeheadline{\vbox to 0pt{\vskip-22.5pt
     \fullline{\vbox to8.5pt{}\the\headline}\vss}\nointerlineskip}
\gdef\makehedlinebb{\vbox to 0pt{\vskip-22.5pt
     \fullline{\vbox to8.5pt{}\copy\hedlinebb\hfil
     \line{\hfill\the\headline\hfill}}\vss} \nointerlineskip}
\gdef\makefootline{\baselineskip=24pt \fullline{\the\footline}}
\gdef\makefotlinebb{\baselineskip=24pt
    \fullline{\copy\fotlinebb\hfil\line{\hfill\the\footline\hfill}}}
\gdef\doubleformat{\shipout\vbox{\Landspec\makehedlinebb
     \fullline{\box\leftcolumn\hfil\columnbox}\makefotlinebb}
     \advancepageno}
\gdef\columnbox{\leftline{\pagebody}}
\gdef\line#1{\hbox to\hsize{\hskip\leftskip#1\hskip\rightskip}}
\gdef\fullline#1{\hbox to\fullhsize{\hskip\leftskip{#1}%
\hskip\rightskip}}
\gdef\footnote#1{\let\@sf=\empty
         \ifhmode\edef\#sf{\spacefactor=\the\spacefactor}\/\fi
         #1\@sf\vfootnote{#1}}
\gdef\vfootnote#1{\insert\footins\bgroup
         \ifnum\dimnota=1  \eightpoint\fi
         \ifnum\dimnota=2  \ninepoint\fi
         \ifnum\dimnota=0  \tenpoint\fi
         \interlinepenalty=\interfootnotelinepenalty
         \splittopskip=\ht\strutbox
         \splitmaxdepth=\dp\strutbox \floatingpenalty=20000
         \leftskip=\oldssposta \rightskip=\olddsposta
         \spaceskip=0pt \xspaceskip=0pt
         \ifnum\sinnota=0   \textindent{#1}\fi
         \ifnum\sinnota=1   \item{#1}\fi
         \footstrut\futurelet\next\fo@t}
\gdef\fo@t{\ifcat\bgroup\noexpand\next \let\next\f@@t
             \else\let\next\f@t\fi \next}
\gdef\f@@t{\bgroup\aftergroup\@foot\let\next}
\gdef\f@t#1{#1\@foot} \gdef\@foot{\strut\egroup}
\gdef\footstrut{\vbox to\splittopskip{}}
\skip\footins=\bigskipamount
\count\footins=1000  \dimen\footins=8in
\catcode`@=12
\tenpoint
\ifnum\unoduecol=1 \hsize=\tothsize   \fullhsize=\tothsize \fi
\ifnum\unoduecol=2 \hsize=\collhsize  \fullhsize=\tothsize \fi
\global\let\lrcol=L      \ifnum\unoduecol=1
\output{\plainoutput{\ifnum\tipbnota=2 \clearnmbnota\fi}} \fi
\ifnum\unoduecol=2 \output{\if L\lrcol
     \global\setbox\leftcolumn=\columnbox
     \global\setbox\fotlinebb=\line{\hfill\the\footline\hfill}
     \global\setbox\hedlinebb=\line{\hfill\the\headline\hfill}
     \advancepageno  \global\let\lrcol=R
     \else  \doubleformat \global\let\lrcol=L \fi
     \ifnum\outputpenalty>-20000 \else\dosupereject\fi
     \ifnum\tipbnota=2\clearnmbnota\fi }\fi
\def\ifdoublepage{\ifnum\unoduecol=2 }
\gdef\yespagenumbers{\footline={\hss\tenrm\folio\hss}}
\gdef\ciao{ \ifnum\fdefcontre=1 \endfdef\fi
     \par\vfill\supereject \ifnum\unoduecol=2
     \if R\lrcol  \headline={}\nopagenumbers\null\vfill\eject
     \fi\fi \end}

\newskip\olddsposta \newskip\oldssposta
\global\oldssposta=\leftskip \global\olddsposta=\rightskip

\def\filldots{\leaders\hbox to 1em{\hss.\hss}\hfill}
\def\inquadrb#1 {\vbox {\hrule  \hbox{\vrule \vbox {\vskip .2cm
    \hbox {\ #1\ } \vskip .2cm } \vrule  }  \hrule} }
 \def\newline{\hfil\break}
\def\jump{\vskip\baselineskip} \newskip\iinnffrr
\def\sjump{\iinnffrr=\baselineskip
          \divide\iinnffrr by 2 \vskip\iinnffrr}
\def\bjump{\vskip\baselineskip \vskip\baselineskip}
\newcount\nmbnota  \def\clearnmbnota{\global\nmbnota=0}
\newcount\tipbnota \def\letterfootnote{\global\tipbnota=1}

\def\note#1{\global\advance\nmbnota by 1 \ifnum\tipbnota=1
    \footnote{$^{\rm\nttlett}$}{#1} \else {\ifnum\tipbnota=2
    \footnote{$^{\nttsymb}$}{#1}
    \else\footnote{$^{\the\nmbnota}$}{#1}\fi}\fi}
\def\nttlett{\ifcase\nmbnota \or a\or b\or c\or d\or e\or f\or
g\or h\or i\or j\or k\or l\or m\or n\or o\or p\or q\or r\or
s\or t\or u\or v\or w\or y\or x\or z\fi}
\def\nttsymb{\ifcase\nmbnota \or\dag\or\sharp\or\ddag\or\star\or
\natural\or\flat\or\clubsuit\or\diamondsuit\or\heartsuit
\or\spadesuit\fi}   \clearnmbnota
\def\numberfootnote{\global\tipbnota=0} \numberfootnote
\def\setnote#1{\expandafter\xdef\csname#1\endcsname{
\ifnum\tipbnota=1 {\rm\nttlett} \else {\ifnum\tipbnota=2
{\nttsymb} \else \the\nmbnota\fi}\fi} }
\newcount\nbmfig  \def\clearnbmfig{\global\nbmfig=0}
\gdef\figure{\global\advance\nbmfig by 1
      {\rm fig. \the\nbmfig}}   \clearnbmfig
\def\setfig#1{\expandafter\xdef\csname#1\endcsname{fig. \the\nbmfig}}
 \def\endformula{\eqno\numero $$}
 \def\efr{\endformula}
\newcount\frmcount \def\clearfrmcount{\global\frmcount=0}
\def\numero{\global\advance\frmcount by 1   \ifnum\indappcount=0
  {\ifnum\cpcount <1 {\hbox{\rm (\the\frmcount )}}  \else
  {\hbox{\rm (\the\cpcount .\the\frmcount )}} \fi}  \else
  {\hbox{\rm (\applett .\the\frmcount )}} \fi}
\def\nameformula#1{\global\advance\frmcount by 1%
\ifnum\draftnum=0  {\ifnum\indappcount=0%
{\ifnum\cpcount<1\xdef\spzzttrra{(\the\frmcount )}%
\else\xdef\spzzttrra{(\the\cpcount .\the\frmcount )}\fi}%
\else\xdef\spzzttrra{(\applett .\the\frmcount )}\fi}%
\else\xdef\spzzttrra{(#1)}\fi%
\expandafter\xdef\csname#1\endcsname{\spzzttrra}
\eqno \hbox{\rm\spzzttrra} $$}
\def\nfr{\nameformula}    
\def\nameali#1{\global\advance\frmcount by 1%
\ifnum\draftnum=0  {\ifnum\indappcount=0%
{\ifnum\cpcount<1\xdef\spzzttrra{(\the\frmcount )}%
\else\xdef\spzzttrra{(\the\cpcount .\the\frmcount )}\fi}%
\else\xdef\spzzttrra{(\applett .\the\frmcount )}\fi}%
\else\xdef\spzzttrra{(#1)}\fi%
\expandafter\xdef\csname#1\endcsname{\spzzttrra}
  \hbox{\rm\spzzttrra} }      \clearfrmcount
\newcount\cpcount \def\clearcpcount{\global\cpcount=0}
\newcount\subcpcount \def\clearsubcpcount{\global\subcpcount=0}
\newcount\appcount \def\clearappcount{\global\appcount=0}
\newcount\indappcount \def\clearindappcount{\indappcount=0}
\newcount\sottoparcount 

\def\applett{\ifcase\appcount  \or {A}\or {B}\or {C}\or
{D}\or {E}\or {F}\or {G}\or {H}\or {I}\or {J}\or {K}\or {L}\or
{M}\or {N}\or {O}\or {P}\or {Q}\or {R}\or {S}\or {T}\or {U}\or
{V}\or {W}\or {X}\or {Y}\or {Z}\fi    \ifnum\appcount<0
\immediate\write16 {Panda ERROR - Appendix: counter "appcount"
out of range}\fi  \ifnum\appcount>26  \immediate\write16 {Panda
ERROR - Appendix: counter "appcount" out of range}\fi}
\clearappcount  \clearindappcount \newcount\connttrre
\def\clearconnttrre{\global\connttrre=0} \newcount\countref
\def\clearcountref{\global\countref=0} \clearcountref
\def\chapter#1{\global\advance\cpcount by 1 \clearfrmcount
                 \goodbreak\null\vbox{\jump\nobreak
                 \clearsubcpcount\clearindappcount
                 \itemitem{\ttaarr\the\cpcount .\qquad}{\ttaarr #1}
                 \par\nobreak\jump\sjump}\nobreak}
\def\section#1{\global\advance\subcpcount by 1 \goodbreak\null
               \vbox{\sjump\nobreak\ifnum\indappcount=0
                 {\ifnum\cpcount=0 {\itemitem{\ppaarr
               .\the\subcpcount\quad\enskip\ }{\ppaarr #1}\par} \else
                 {\itemitem{\ppaarr\the\cpcount .\the\subcpcount\quad
                  \enskip\ }{\ppaarr #1} \par}  \fi}
                \else{\itemitem{\ppaarr\applett .\the\subcpcount\quad
                 \enskip\ }{\ppaarr #1}\par}\fi\nobreak\jump}\nobreak}
\clearsubcpcount
\def\appendix#1{\global\advance\appcount by 1 \clearfrmcount
                  \goodbreak\null\vbox{\jump\nobreak
                  \global\advance\indappcount by 1 \clearsubcpcount
          \itemitem{ }{\hskip-40pt\ttaarr #1}
%                  \itemitem{\ttaarr App.\applett\ }{\ttaarr #1}
             \nobreak\jump\sjump}\nobreak}
\clearappcount \clearindappcount
\def\references{\goodbreak\null\vbox{\jump\nobreak
   \noindent{\ttaarr References} \nobreak\jump\sjump}\nobreak}
%   \itemitem{}{\ttaarr References} \nobreak\jump\sjump}\nobreak}

\clearcpcount\clearcountref

\def\setchap#1{\ifnum\indappcount=0{\ifnum\subcpcount=0%
\xdef\spzzttrra{\the\cpcount}%
\else\xdef\spzzttrra{\the\cpcount .\the\subcpcount}\fi}
\else{\ifnum\subcpcount=0 \xdef\spzzttrra{\applett}%
\else\xdef\spzzttrra{\applett .\the\subcpcount}\fi}\fi
\expandafter\xdef\csname#1\endcsname{\spzzttrra}}
\newcount\draftnum \newcount\ppora   \newcount\ppminuti
\global\ppora=\time   \global\ppminuti=\time
\global\divide\ppora by 60  \draftnum=\ppora
\multiply\draftnum by 60    \global\advance\ppminuti by -\draftnum
\def\droggi{\number\day /\number\month /\number\year\ \the\ppora
:\the\ppminuti}     \global\draftnum=0
\def\draftcomment#1{\ifnum\draftnum=0 \relax \else
{\ {\bf ***}\ #1\ {\bf ***}\ }\fi} 
%
%     Maximum number of references = 200
%     boxes 50 -> 250 reserved for references
%
\catcode`@=11
\gdef\Ref#1{\expandafter\ifx\csname @rrxx@#1\endcsname\relax%
{\global\advance\countref by 1    \ifnum\countref>200
\immediate\write16 {Panda ERROR - Ref: maximum number of references
exceeded}  \expandafter\xdef\csname @rrxx@#1\endcsname{0}\else
\expandafter\xdef\csname @rrxx@#1\endcsname{\the\countref}\fi}\fi
\ifnum\draftnum=0 \csname @rrxx@#1\endcsname \else#1\fi}
\gdef\beginref{\ifnum\draftnum=0  \gdef\Rref{\fairef}
\gdef\endref{\scriviref} \else\relax\fi
\ifx\risposta\mplarisposta \ninepoint \fi
\parskip 2pt plus.2pt \baselineskip=12pt}
\def\Reflab#1{[#1]} \gdef\Rref#1#2{\item{\Reflab{#1}}{#2}}
\gdef\endref{\relax}  \newcount\conttemp
\gdef\fairef#1#2{\expandafter\ifx\csname @rrxx@#1\endcsname\relax
{\global\conttemp=0 \immediate\write16 {Panda ERROR - Ref: reference
[#1] undefined}} \else
{\global\conttemp=\csname @rrxx@#1\endcsname } \fi
\global\advance\conttemp by 50  \global\setbox\conttemp=\hbox{#2} }
\gdef\scriviref{\clearconnttrre\conttemp=50
\loop\ifnum\connttrre<\countref \advance\conttemp by 1
\advance\connttrre by 1
\item{\Reflab{\the\connttrre}}{\unhcopy\conttemp} \repeat}
\clearcountref \clearconnttrre
\catcode`@=12
\ifx\risposta\mplarisposta \def\Reflab#1{#1.} \letterfootnote \fi

\def\slashchar#1{\setbox0=\hbox{$#1$} \dimen0=\wd0
     \setbox1=\hbox{/} \dimen1=\wd1 \ifdim\dimen0>\dimen1
      \rlap{\hbox to \dimen0{\hfil/\hfil}} #1 \else
      \rlap{\hbox to \dimen1{\hfil$#1$\hfil}} / \fi}
\ifx\oldchi\undefined \let\oldchi=\chi
  \def\cchi{{\raise 1pt\hbox{$\oldchi$}}} \let\chi=\cchi \fi

\def\frac#1#2{{\textstyle{#1 \over #2}}}

\def\half{\ifinner {\scriptstyle {1 \over 2}}\else {1 \over 2} \fi}
\def\bra#1{\langle#1\vert}  \def\ket#1{\vert#1\rangle}

\def\simge{\rlap{\raise 2pt \hbox{$>$}}{\lower 2pt \hbox{$\sim$}}}
\def\simle{\rlap{\raise 2pt \hbox{$<$}}{\lower 2pt \hbox{$\sim$}}}

\def\buildchar#1#2#3{{\null\!\mathop{#1}\limits^{#2}_{#3}\!\null}}
\def\overcirc#1{\buildchar{#1}{\circ}{}}

\def\vbig#1#2{{\vbigd@men=#2\divide\vbigd@men by 2%
\hbox{$\left#1\vbox to \vbigd@men{}\right.\n@space$}}}

%
%--------------------------------------------------------------------
%
\newcount\fdefcontre \newcount\fdefcount \newcount\indcount
\newread\filefdef  \newread\fileftmp  \newwrite\filefdef
\newwrite\fileftmp     \def\strip#1*.A {#1}
\def\futuredef#1{\beginfdef
\expandafter\ifx\csname#1\endcsname\relax%
{\immediate\write\fileftmp {#1*.A}
\immediate\write16 {Panda Warning - fdef: macro "#1" on page
\the\pageno \space undefined}
\ifnum\draftnum=0 \expandafter\xdef\csname#1\endcsname{(?)}
\else \expandafter\xdef\csname#1\endcsname{(#1)} \fi
\global\advance\fdefcount by 1}\fi   \csname#1\endcsname}

\def\beginfdef{\ifnum\fdefcontre=0
\immediate\openin\filefdef \jobname.fdef
\immediate\openout\fileftmp \jobname.ftmp
\global\fdefcontre=1  \ifeof\filefdef \immediate\write16 {Panda
WARNING - fdef: file \jobname.fdef not found, run TeX again}
\else \immediate\read\filefdef to\spzzttrra
\global\advance\fdefcount by \spzzttrra
\indcount=0      \loop\ifnum\indcount<\fdefcount
\advance\indcount by 1   \immediate\read\filefdef to\spezttrra
\immediate\read\filefdef to\sppzttrra
\edef\spzzttrra{\expandafter\strip\spezttrra}
\immediate\write\fileftmp {\spzzttrra *.A}
\expandafter\xdef\csname\spzzttrra\endcsname{\sppzttrra}
\repeat \fi \immediate\closein\filefdef \fi}
\def\endfdef{\immediate\closeout\fileftmp   \ifnum\fdefcount>0
\immediate\openin\fileftmp \jobname.ftmp
\immediate\openout\filefdef \jobname.fdef
\immediate\write\filefdef {\the\fdefcount}   \indcount=0
\loop\ifnum\indcount<\fdefcount    \advance\indcount by 1
\immediate\read\fileftmp to\spezttrra
\edef\spzzttrra{\expandafter\strip\spezttrra}
\immediate\write\filefdef{\spzzttrra *.A}
\edef\spezttrra{\string{\csname\spzzttrra\endcsname\string}}
\iwritel\filefdef{\spezttrra}
\repeat  \immediate\closein\fileftmp \immediate\closeout\filefdef
\immediate\write16 {Panda Warning - fdef: Label(s) may have changed,
re-run TeX to get them right}\fi}
\def\iwritel#1#2{\newlinechar=-1
{\newlinechar=`\ \immediate\write#1{#2}}\newlinechar=-1}
\global\fdefcontre=0 \global\fdefcount=0 \global\indcount=0
%
%--------------------------------------------------------------------
%
\null
%
%--------------------------------------------------------------------
%
%                             THE    END
%
%--------------------------------------------------------------------
%\input panda
%\draftmode{Tau-functions.....}
\loadamsmath
\loadeuler
%
%%%%%%%%%%%%%%%%%%%%%%%%%%%%%
%
% SOME SPECIAL CHARACTERS
%
%%%%%%%%%%%%%%%%%%%%%%%%%%%%%
%
\def\ggg{{\eufm g}}

\def\ie{{\it i.e.\/}}
\def\eg{{\it e.g.\/}}
\def\gg{{\>\widehat{g}\>}}
\def\GG{{\>\widehat{G}\>}}
\def\ss{{\>\widehat{s}\>}}
\def\s{{\bf s}}
\def\sw{{\bf s}^w}
\def\Heis{{\cal H}[w]}
\def\ad{{\rm ad\>}}
\def\Ker{{\rm Ker\/}}
\def\Im{{\rm Im\/}}
\def\Pos{{\rm P}_{\geq0[\sw]}}

\def\cl{{\cal L}}
\def\pa{\partial}
%%%%%%%%%%%%%%%%%
\mathchardef\bphi="731E
\mathchardef\balpha="710B
\mathchardef\bomega="7121
\def\alb{{\bfmath\balpha}}
\def\pp{{\bfmath\bphi}}

\def\Hb{{\bfmath H}}
\pageno=0
\nopagenumbers{\baselineskip=12pt
\line{\hfill US-FT/25-96}
\line{\hfill IFT-P.016/96}
\line{\hfill\tt hep-th/9606066}
\line{\hfill June 1996}\jump
\ifdoublepage \bjump\bjump\bjump\else\jump\vfill\fi
\centerline{{\capstwo Tau-functions and Dressing Transformations}}
\sjump
\centerline{{\capstwo for Zero-Curvature Affine Integrable Equations}} 
\bjump
\centerline{{\scaps Luiz~A.
Ferreira{$^*$},
J.~Luis Miramontes$^{\dagger}$ and  Joaqu\'\i n S\'anchez
Guill\'en$^{\dagger}$}}\jump

\centerline{$^*${\sl Instituto de F\' \i sica Te\'orica, IFT/UNESP}}
\centerline{\sl Universidade Estadual Paulista}
\centerline{\sl Rua Pamplona 145}
\centerline{\sl 01405-900 S\~ao Paulo - SP, Brazil}\jump

\centerline{$^{\dagger}${\sl Departamento de F\'\i sica de Part\'\i culas}}
\centerline{\sl Facultad de F\'\i sica}
\centerline{\sl Universidad de Santiago}
\centerline{\sl E-15706 Santiago de Compostela, Spain}\jump

\centerline{{\tt laf@axp.ift.unesp.br},~~ {\tt miramont@fpaxp1.usc.es},~~ and~~
{\tt joaquin@fpaxp1.usc.es}}\bjump\jump

\ifdoublepage
\vfill
{\noindent
\line{June 1996\hfill}}
\eject\null\vfill\fi
\centerline{\capsone ABSTRACT}\jump

\noindent
The solutions of a large class of hierarchies of zero-curvature equations that 
includes Toda and KdV type hierarchies are investigated. All these
hierarchies are constructed from affine (twisted or untwisted) Kac-Moody
algebras~$\ggg$. Their common feature is that they have some special ``vacuum
solutions'' corresponding to Lax operators lying in some abelian (up to the
central term) subalgebra of~$\ggg$; in some interesting cases such subalgebras
are of the Heisenberg type. Using the dressing transformation method, the
solutions in the orbit of those vacuum solutions are constructed in a
uniform way. Then, the generalized tau-functions for those hierarchies are
defined as an alternative set of variables corresponding to certain matrix
elements evaluated in the integrable highest-weight representations of~$\ggg$.
Such definition of tau-functions applies for any level of the representation,
and it is independent of its realization (vertex operator or not).  The
particular important cases of generalized mKdV and KdV hierarchies as well as
the abelian and non abelian affine Toda theories are discussed in detail.   

\sjump\vfill
\ifdoublepage \else
\noindent
\line{June 1996\hfill}\fi
\eject}
\yespagenumbers\pageno=1
\footline={\hss\tenrm-- \folio\ --\hss}
%\doublespaced
%
%%%%%%%%%%%%%%%%%%%---------------!!!!!!!!!---------------%%%%%%%%%%%%%%%%%%%

\chapter{Introduction.}

In this paper we shall be concerned with the generalization of the Hirota
method of constructing the solutions of hierarchies of non-linear
integrable models. In particular, we shall study such connections through a 
large and important class of solutions which can be constructed in a uniform 
way using the underlying structure of affine Kac-Moody 
algebras of those  hierarchies.  

Out of the different available methods for solving integrable partial 
differential equations, the Hirota method has proved to be
particularly useful. This method started with the work of R.~Hirota~[\Ref{HIR}],
who discovered a way to construct various types of explicit solutions to the
equations, and, in particular, their multiple soliton solutions.
The idea is to find a new set of variables, called the ``tau-functions'', which
then satisfy simpler ---originally bilinear--- equations known as Hirota
equations. For instance, the tau-function of the Korteweg-de Vries equation
(KdV), 
$ \partial_t u = \partial_{x}^3 u + 6 u\>\partial_x u $, 
is related to the original variable by the celebrated formula 
$$
u =2\> \partial_{x}^2\ln \tau \>,
\nfr{kdvtau}
Such tau-function satisfies a bilinear Hirota equation~[\Ref{HKDV}], 
and the exact multi-soliton solutions are found by considering truncated 
 series
expansions of $\tau$ in some arbitrary parameter $\epsilon$, \eg, $\tau= 1 +
\epsilon\>\tau^{(1)}+ \cdots \epsilon^n \> \tau^{(n)}$.  

More recently, the Hirota method has been used to obtain the  multi-soliton
solutions of affine (abelian) Toda
equations~[\Ref{TIMSOL},\Ref{HIRL},\Ref{mm},\Ref{MAX}] (the method was
originally applied to the sine-Gordon equation in ref.~[\Ref{HSG}]). The
success of the method depends crucially on the choice of the change of
variables between the Toda fields $\pp$ and the Hirota's tau functions
$\tau_i$, namely
$$
\pp = -\sum_{i=0}^r {2\over \alb_{i}^2}\> \alb_i\> \ln
\tau_i\>,
\nfr{TodaTau}
where $\alb_i$ are the simple roots of the associated affine untwisted 
Kac-Moody  algebra. 

A priori, the origin of formulae like $ u =2\> \partial_{x}^2\ln \tau$ or
\TodaTau\ seems quite mysterious and unmotivated. Nevertheless, for a large
class of integrable equations, they have a remarkable group theoretical
interpretation within the, so called, tau-function approach pioneered by the
japanese school~(see for example~[\Ref{TAUJAP}]). Actually, this approach
manifests the deep underlying connection of the integrable hierarchies of
partial differential equations with affine Kac-Moody algebras; a connection
that is also apparent in the seminal work of Drinfel'd and Sokolov~[\Ref{DS}],
where integrable hierarchies of equations are constructed in zero-curvature
form.

The tau-function approach has been largely clarified by the
work of Wilson~[\Ref{WILa},\Ref{WILb}] and of Kac and Wakimoto~[\Ref{KW}]. In
this latter reference, the authors construct hierarchies of integrable equations
directly in Hirota form associated to vertex operator representations of
Kac-Moody algebras; then, the tau-functions describe the orbit of
the highest-weight vector of the representation under the corresponding
Kac-Moody group. On the other hand, the work of Wilson provides the group
theoretical interpretation of the change of variables between the tau-functions
and the natural variables in the zero curvature approach for several well known
integrable equations like KdV and modified KdV~[\Ref{WILa}], and non-linear
Schr\"odinger~[\Ref{WILb}] (see also~[\Ref{BK}]). In these articles, the change
of variables is obtained by using a particular version of the well known
dressing transformations of Zakharov and Shabat~[\Ref{ZS}]. 

Using Wilson's ideas, the connection between the generalized Hirota
equations of Kac and Wakimoto and the zero-curvature equations
of~[\Ref{GEN}] has been established in ref.~[\Ref{TAUTIM}]. 
It is worth noticing that the class of integrable equations of~[\Ref{GEN}]
is large enough to include practically all the generalizations of the
Drinfel'd-Sokolov construction considered so far in the literature, and,
therefore, it is desirable to have the tau-function description of all those
integrable hierarchies of integrable equations. The reason why the results
of~[\Ref{TAUTIM}] do not apply for all the integrable hierarchies
of~[\Ref{GEN}] is that the generalized Hirota equations of~[\Ref{KW}] are
constructed in terms of level-one vertex operator representations of simply
laced affine Kac-Moody algebras, while the integrable hierarchies
of~[\Ref{GEN}] require a definition of the tau-functions in terms of arbitrary
highest-weight representations. 

Another important restriction in the results of~[\Ref{TAUTIM}] is that they do
not include the important class of integrable equations known as generalized
Toda equations; \eg, they do not explain the change of variables~\TodaTau.
Nevertheless, inspired by the results of~[\Ref{TAUTIM}], a definition for the
tau-functions of the Toda equations has been proposed in~[\Ref{LUIZ}].

The aim of this paper is to generalize the results of~[\Ref{TAUTIM}]
and~[\Ref{LUIZ}] in order to clarify the definition and relevance of
tau-functions for a large class of integrable equations including both the
integrable hierarchies of~[\Ref{GEN}] and the non-abelian
generalizations of the Toda equation. In our approach, the central role will
be played by the dressing transformations in the manner described by
Wilson~[\Ref{WILa},\Ref{WILb}]. This way, we will construct explicit solutions 
of
certain non-linear integrable equations by dressing some  ``vacuum solutions''.
Actually, we will recognize the relevant equations by inspecting the properties
of the resulting solutions. This is reminiscent of the tau-function approach
of~[\Ref{KW}], where the tau-functions are defined as the elements of the orbit
of a highest-weight under the Kac-Moody group, and the generalized Hirota
equations are just the equations characterizing those orbits; thus, the
solutions and the equations are obtained simultaneously. In contrast, with our
method we do not expect to produce all the solutions of the resulting equations,
but only a subset that is conjectured to include the multi-soliton solutions. 

The paper is organized as follows. In section~2 we describe the type of 
 hierarchies we are going to consider, discuss their vacuum solutions 
and construct 
solutions using the dressing transformation method. In section~3 we define  
the tau-functions for all these hierarchies using integrable highest-weight 
representations of affine Kac-Moody algebras and generalizing some results 
known for level-one vertex operator representations. In section~4 we
specialize our results to the generalized mKdV (and KdV)
hierarchies of~[\Ref{GEN}], and to the abelian and non abelian affine Toda
theories.  Conclusions are presented in section~5, and we also provide an
appendix with our conventions about Kac-Moody algebras and their integrable
highest-weight representations. 

\chapter{Vacuum solutions and dressing transformations.}

Non-linear integrable hierarchies of equations are most conveniently
discussed by associating them with a system of first-order
differential equations
$$
\cl_N \Psi = 0\>,
\nfr{LinProb}
where $\cl_N $ are Lax operators of the form
$$
\cl_N  \equiv {\pa \, \over \pa t_N} - A_N 
\nfr{LaxGen}
and the variables $t_N$ are the various ``times'' of the hierarchy. Then, the
equivalent zero-curvature formulation is obtained through the integrability 
conditions of the associated linear problem~\LinProb,
$$
[{\cal L}_N\>, \>{\cal L}_M]=0\>. 
\nfr{ZeroCurv}
An equivalent way to express the relation between the solutions of the
zero-curvature equations and of the associated linear problem is
$$
A_N = {\partial \Psi\over \partial t_{N}}\>
\Psi^{-1}\>.
\nfr{psidef}

The class of integrable hierarchies of zero-curvature equations that will be
studied here is constructed from graded Kac-Moody
algebras in the following way (we have briefly summarized our conventions
concerning Kac-Moody algebras in the appendix). Consider a complex affine
Kac-Moody algebra 
$\ggg = \gg \oplus {\Bbb C}\> d$ of rank $r$, and an integer gradation of its
derived algebra $\gg$ labelled by a vector $\s=(s_0,s_1,\ldots,s_r)$ of
$r+1$ non-negative co-prime integers such that 
$$
\gg= \bigoplus_{i\in{\Bbb Z}} \gg_i(\s)\> \quad {\rm
and}\quad [\gg_i(\s),\gg_j(\s)] \subseteq \gg_{i+j}(\s)\>. 
\nfr{Gradation}

We have in mind basically two types of integrable systems. The first one
corresponds to the Generalized Drinfel'd-Sokolov Hierarchies considered
in~[\Ref{GEN},\Ref{TAUTIM}], which are generalizations of the KdV type
hierarchies studied in~[\Ref{DS}]. In particular, and using the parlance of
the original references, we will be interested in the generalized mKdV
hierarchies, whose construction can be summarised as follows
(see~[\Ref{GEN}] and, especially,~[\Ref{TAUTIM}] for details). Given an
integer gradation
$\s$ of
$\gg$ and a semisimple constant element $E_l$ of grade $l$ with respect to
$\s$, one defines the Lax  operator 
$$
L \equiv \pa_x + E_l + A\>,
\nfr{xlax}
where the components of $A$ are the fields of the
hierarchy.~\note{In~[\Ref{TAUTIM}] it was shown that the component of $A$ along
the central term of~$\ggg$ should not be considered as an actual degree of
freedom of the hierarchy. This is the reason why these hierarchies can be
equivalently formulated both in terms of affine Kac-Moody algebras or of the
corresponding loop algebras (see section 4.1 for more details about this).} They
are functions of
$x$ and of the other times of the hierarchy taking values in the subspaces  of
$\gg$ with grades ranging from $0$ to $l-1$. For each element in the centre of
$\Ker(\ad E_l )$ with positive $\s$-grade $N$, one constructs a local
functional of those fields,
$B_N$, whose components take values in the subspaces
$\gg_0(\s), \ldots, \gg_N(\s)$. Then, $B_N$ defines the flow equation
$$
{\pa L\over \pa t_N}\> =\>  \bigl[ B_N \> , \> L \bigr]\>,
\nfr{flow}
and the resulting Lax operators $\cl_N = {\pa /\pa t_N}  - B_N$ commute among 
themselves~[\Ref{GEN}].

The second type of integrable systems corresponds to the non-abelian affine
Toda theories~[\Ref{LS},\Ref{LUIZ},\Ref{SAVGERV},\Ref{UNDER}]. Given the
integer gradation $\s$ of $\gg$, one chooses two constant elements
$E_{\pm l}$  in $\gg_{\pm l}(\s)$ and introduces two Lax operators
$$
L_+ = \pa_+ - B F^+ B^{-1} \> , \qquad 
L_- = \pa_- - \pa_- B \> B^{-1} + F^-\>.
\nfr{todalax}
The field $B$ is a function of $x_\pm$ taking values in the group obtained
by exponentiating the zero graded subalgebra $\gg_0(\s)$. As for the other
fields, the functions
$F^{\pm}$ can be decomposed as $F^{\pm} = E_{\pm l} +
\sum_{m=1}^{l-1} F^{\pm}_m$, and $F^{+}_m$ and $F^{-}_m$ take values in
$g_{m}(\s)$ and $g_{-m}(\s)$, respectively. Then, the condition $[ L_{+} \> ,
\> L_{-}]=0$ provides the equations-of-motion of the theory, where $\pa_{\pm}$
are the derivatives with respect to the light-cone variables $x_{\pm}$.
The well known abelian affine Toda equations are
recovered with the principal gradation, $\s=(1,1,\ldots,1)$, and $l=1$.
They possess an infinite number of conserved charges in
involution~[\Ref{OLTUR}], and these charges can be used to construct a
hierarchy of integrable models through an infinite number of Lax
operators that commute among themselves~[\Ref{OLTUR2}].
The non-abelian versions of the affine Toda equations are obtained with
generic gradations $\s$ and $F_{m}^\pm =0$~[\Ref{LS},\Ref{LUIZ},\Ref{UNDER}],
while the most general case with $F_{m}^\pm \not=0$ corresponds to the coupling
of the latter systems with (spinor) matter fields~[\Ref{SAVGERV}].

An important common feature of all those hierarchies is that they possess
trivial solutions which will be called ``vacuum solutions''. These 
particular solutions are singled out by the condition that the  Lax operators
evaluated on them lie on some abelian subalgebra of~\ggg, up to central terms.
Then, the dressing  transformation method can be used to generate an orbit of
solutions out of each ``vacuum''. Moreover, it is generally conjectured that
multi-soliton solutions lie  in the resulting orbits. As a bonus, the fact that
we only consider the particular subset of solutions connected with a generic
vacuum  allows one to perform the calculations in a  very general way and,
consequently, our results  apply to a much broader class of hierarchies. 

For a given choice of the Kac-Moody algebra $\ggg$ and the gradation $\s$, let 
us
consider Lax  operators of the form \LaxGen\ where the potentials can be
decomposed as
$$
A_N = \sum_{i=N_{-}}^{N_{+}} A_{N,i} \> , \quad  {\rm where}\quad
A_{N,i}  \in \gg_i(\s)
\nfr{genpot}
$N_{-}$ and $N_{+}$ are non-positive and non-negative integers, 
respectively, and the times $t_N$ are labelled by (positive or
negative) integer numbers. The particular form of these potentials will be
constrained only by the condition that the corresponding hierarchy admits vacuum
solutions where they take the form  
$$
A_N^{({\rm vac})}\> =\>  \sum_{i=N_{-}}^{N_{+}} c_N^i b_i+ \rho_N (t)\> c
\>\equiv  \> \varepsilon_N + \rho_N (t)\> c\>. 
\nfr{vacpot}
In this equation, $c$ is the central element of $\gg$, and  $b_i\in \gg_i(\s)$ 
are the generators of a subalgebra $\ss $ of $\gg$ defined by
$$
\ss = \{b_i \in\gg_i(\s)\>,\; i\in E\subset {\Bbb Z}\bigm| [b_i,b_j] =
i \>\beta_i\>c\> \delta_{i+j,0}\}\>,
\nfr{Commute}
where  $\beta_i$ are arbitrary (vanishing or non-vanishing) complex numbers
such that $\beta_{-i} = \beta_i$, and
$E$ is some set of integers numbers. Moreover, $c_N^i$ are also arbitrary
numbers, and $\rho_N (t)$ are $\Bbb C$-functions of the times $t_N$ that satisfy
the equations
$$
{\pa\> \rho_N(t)\over \pa\> t_M} \> -\> {\pa\> \rho_M(t)\over \pa\> t_N}\> = \>
\sum_{i}\> i\> \beta_i\> c_M^i\> c_N^{-i}\>. 
\nfr{ZeroTriv} 

These vacuum potentials correspond to the solution of the associated linear
problem given by the group  element~\psidef\ 
$$
\Psi^{({\rm vac})} = \exp \Bigl( \sum_{N} \varepsilon_N t_N
\> +\> \gamma (t) \> c \Bigr)
\nfr{vacelem}
where the numeric function $\gamma (t)$ is a solution of the equations
$$
{\pa \gamma (t) \over \pa t_N} = \rho_N (t) + {1\over 2}\sum_{M,i}
i\>\beta_i\> c_N^i\> c_M^{-i}\>t_M\>. 
\nfr{gamma}

In terms of the associated linear problem, one can define an important set of
transformations called ``dressing transformations'', which take known solutions
of the hierarchy to new solutions. Regarding the structure of the integrable
hierarchies, these transformations have a deep meaning and, in fact,
the group of dressing transformations can be viewed as the classical precursor
of the quantum group symmetries~[\Ref{BABELON}]. Denote by $\GG_-(\s)$, 
 $\GG_+(\s)$, and $\GG_0(\s)$ the subgroups of the Kac-Moody group $\GG$ formed
by exponentiating the subalgebras $\gg_{<0}(\s ) \equiv \bigoplus_{i<0}
\gg_i(\s)$, $\gg_{>0}(\s)\equiv \bigoplus_{i>0} \gg_i(\s)$, and $\gg_0(\s)$,
respectively. According to Wilson~[\Ref{WILa},\Ref{WILb}], the dressing
transformations can be described in the following way. Consider a solution
$\Psi$ of the linear problem~\LinProb, and let 
$h= h_-\> h_0\> h_+$ be a constant element in the ``big cell'' of $\GG$, {\it
i.e.\/}, in the subset $\GG_-(\s)\> \GG_0(\s)\> \GG_+(\s)$ of $\GG$, such that 
$$
\Psi\> h\> \Psi^{-1} = (\Psi\> h\> \Psi^{-1})_-\> (\Psi\> h\>
\Psi^{-1})_0\> (\Psi\> h\> \Psi^{-1})_+\>.
\nfr{Factor}
Notice that these conditions are equivalent to say that both $h$ and $\Psi\> 
h\> \Psi^{-1}$ admit a generalized Gauss decomposition with respect to the
gradation $\s\>$. Then 
$$
\eqalign{
\Psi^h \> & = \> [(\Psi\> h\> \Psi^{-1})_-]^{-1}\> \Psi\cr
&  =\> (\Psi\> h\> \Psi^{-1})_0\> (\Psi\> h\> \Psi^{-1})_+ \>\Psi\> 
h^{-1}\cr}
\nfr{Dressing}
is another solution of the linear problem. In order to prove it, introduce the
notation $g_{0,\pm} \equiv (\Psi\> h\> \Psi^{-1})_{0,\pm}$ and $\partial_N
\equiv \partial/\partial t_N$, and consider
$$
\eqalign{
\partial_N \Psi^h\> {\Psi^h}^{-1} & = 
- g_-^{-1}\> \partial_N g_-   + 
g_-^{-1} (\partial_N \Psi \> \Psi^{-1}) g_- \cr
& = \partial_N g_0 \> g_0^{-1} + 
g_0 \partial_N g_+ \> g_+^{-1} g_0^{-1} + 
g_0 g_+ (\partial_N \Psi\> \Psi^{-1}) g_+^{-1} g_0^{-1}}
\efr
Then, the first identity implies that $\partial_N \Psi^h\> {\Psi^h}^{-1} \in 
\bigoplus_{i\leq N_+}\gg_{i}(\s)$, and the second that $\partial_N \Psi^h\> 
{\Psi^h}^{-1} 
\in \bigoplus_{i\geq N_-}\gg_{i}(\s)$. Consequently
$$
A_{N}^h = {\partial \Psi^h\over \partial t_N}\> {\Psi^h}^{-1}
\in \bigoplus_{i=N_-}^{N_+}\gg_{i}(\s)\>,
\nfr{DressZero}
and, taking into account~\genpot, it is a solution of the hierarchy of
zero-curvature equations.~\note{If the fields of the hierarchy are such that
$A_{N,i}$ does not span the whole subspace $\gg_{i}(\s)$ then we have to
impose further constraints on the group elements performing the dressing
transformation (see section~4.2).}

For any $h$ lying in the big cell of $\GG$, the transformation 
$$
{\cal D}_h : \Psi\> \mapsto\> \Psi^h\>, \quad{\rm or}\quad A_{N} \mapsto
A_{N}^h\>,
\nfr{DresTr} 
is called a dressing transformation, and an important property is that
their composition law follows just from the composition law of
$\GG$, \ie, $ {\cal D}_g \circ {\cal D}_h = {\cal D}_{gh}$.~\note{The dressing
transformations of~[\Ref{BABELON}] are defined in a different way through ${\cal
D}^{\ast}_h(\Psi) = [(\Psi h\Psi^{-1})_-]^{-1} \Psi h_-$, which leads to a
different composition law characterised by a classical r-matrix. Nevertheless,
this definition induces exactly the same transformation $A_{N} \mapsto
A_{N}^h$ among the solutions of the zero-curvature equations.} 

Now, the orbit of the vacuum solution~\vacelem\ under the group of dressing
transformations can be easily constructed using eqs.~\Dressing\ and
\DressZero. For any element $h$ of the big cell of $\GG$, let us define
$$
\eqalignno{
&\Theta^{-1}= (\Psi^{(\rm vac)}\> h\> {\Psi^{(\rm vac)}}^{-1})_-
\>,\qquad B^{-1} = (\Psi^{(\rm vac)}\> h\> {\Psi^{(\rm
vac)}}^{-1})_0\>,\cr  & \Upsilon =(\Psi^{(\rm vac)}\> h\> {\Psi^{(\rm
vac)}}^{-1})_+\>,\quad {\rm and} \quad 
\Omega = B^{-1}\>\Upsilon\>. &\nameali{Teta}\cr }
$$
Then, under the dressing transformation generated by $h$,
$$
\Psi^{(\rm vac)}\> \mapsto\> \Psi^h = \Theta\> \Psi^{(\rm vac)} = \Omega\>
\Psi^{(\rm vac)}\> h^{-1}\>,
\nfr{Simple}
or, equivalently, $A_N^{({\rm vac})}$ becomes
$$
\eqalignno{
A_{N}^h - \rho_N(t)\> c &\> =\> \Theta\> \varepsilon_N \> \Theta^{-1}\> +\> 
\partial_{N}\Theta\>\Theta^{-1}\> \in\> \bigoplus_{i\leq N_+} \gg_{i}(\s)\cr 
&\> =\> \Omega\> \varepsilon_N \> \Omega^{-1}\> +\> \partial_{N} 
\Omega\>\Omega^{-1} \>\in\> \bigoplus_{i\geq N_-} \gg_{i}(\s) , & 
\nameali{DPlus}\cr }
$$

Eqs.~\Teta\ and~\DPlus\ summarize the outcome of the dressing transformation
method, which, starting with some vacuum solution~\vacpot, associates a solution
of the zero-curvature equations~\ZeroCurv\ to each constant element $h$ in the
big cell of $\GG$. The construction of this solution involves two
steps. First, the eqs.~\DPlus\ can be
understood as a local change of variables between the components of the
potential $A_N$ and some components of the group elements $\Theta$, $B$ and
$\Upsilon$. 

The second step consists in obtaining the value of the required components
of $\Theta$, $B$ and $\Upsilon$ from eq.~\Teta. This is usually done by
considering matrix elements of the form
$$
\bra{\mu}\> \Theta^{-1}\> B^{-1}\> \Upsilon  \> \ket{\mu^{\prime}} = 
\bra{\mu}\> e^{\sum_{N} \varepsilon_N t_N} \> h \> 
e^{-\sum_{N} \varepsilon_N t_N} \> \ket{\mu^{\prime}}\>,
\nfr{solspec}
where $\ket{\mu}$ and $\ket{\mu^{\prime}}$ are vectors in a given representation 
of $\ggg$. The appropriate set of vectors is specified by the condition that all
the required components of $\Theta$, $B$ and $\Upsilon$ can be expressed in
terms of the resulting matrix elements. It will be show in the next
section that the required matrix elements, considered as functions of the
group element $h$, constitute the generalization of the Hirota's tau-functions
for these hierarchies. Moreover, eq.~\solspec\ is the analogue of the, so 
called,
solitonic specialization of the Leznov-Saveliev solution proposed
in~[\Ref{TUROLA},\Ref{TUROLB},\Ref{SOLSPEC},\Ref{LUIZ},\Ref{SAVGERV}] for the 
affine (abelian
and non-abelian) Toda theories. 

Consider now the common eigenvectors of the adjoint action of the
$\varepsilon_N$'s that specify the vacuum solution~\vacpot. Then, the
important class of multi-soliton solutions is conjectured to
correspond to group elements $h$ which are the product of
exponentials of eigenvectors
$$
h = e^{F_1} \> e^{F_2} \> \ldots e^{F_n} \>, \qquad 
[ \varepsilon_N \> , \> F_k ] = \omega_N^{(k)} \> F_k \> , \quad k=1,2, 
\ldots n\>.
\nfr{eigenb}
In this case, the dependence of the solution upon the times $t_N$ can be made
quite explicit
$$
\bra{\mu}\> \Theta^{-1}\> B^{-1}\> \Upsilon  \> \ket{\mu^{\prime}} = 
\bra{\mu}\> \prod_{k=1}^n \exp (e^{\sum_{N} \omega_{N}^{(k)} t_N} F_k )  
 \> \ket{\mu^{\prime}}\>.
\nfr{solspecb}
The conjecture that multi-soliton solutions are associated with group
elements of the form~\eigenb\ naturally follows from the well known properties 
of
the multi-soliton solutions of affine Toda equations and of hierarchies of the
KdV type, and, in the sine-Gordon theory, it has been explicitly checked in
ref.~[\Ref{BABT}]. Actually, in all these cases, the multi-soliton solutions are
obtained in terms of representations of the ``vertex operator'' type where the
corresponding eigenvectors are nilpotent. Then, for each eigenvector $F_k$ there
exists a positive integer number $m_k$ such that
$(F_k)^{m} \not=0$ only if $m\leq m_k$. This remarkable property simplifies the
form of~\solspecb\ because it implies that
$e^{F_k}\> =\> 1\>+ \> F_k\> + \cdots+ (F_k)^{m_k}/m_k!$, which provides a
group-theoretical justification of Hirota's method.   
 
An interesting feature of the dressing transformations method is the
possibility of relating the solutions of different integrable equations.
Consider two different integrable hierarchies whose vacuum solutions are
compatible, in the sense that the corresponding vacuum Lax operators commute.
Then, one can consider the original integrable equations as the restriction of
a larger hierarchy of equations. Consequently, the solutions obtained through
the group of dressing transformations can also be understood in terms of the
solutions of the larger hierarchy, which implies certain relations among them.
We will show in section~4 that this possibility generalizes the
well knonw relation between the solutions of the mKdV and sine-Gordon
equations.

\chapter{The tau-functions}

According to the discussion in the previous section, the orbits generated by
the group of dressing transformations acting on some vacuum provide
solutions of certain integrable hierarchies of equations. Making contact with
the method of Hirota, the generalized ``tau-functions'' that will be
defined in this section constitute a new set of variables to describe those
solutions. One of the characteristic properties of these variables is that they
substantially simplify the task of constructing multi-soliton
solutions~[\Ref{LUIZ}]. The group-theoretical interpretation of this property
has already been pointed out in the previous section. Tau-functions are given
by certain matrix elements in a appropriate representation of the Kac-Moody
Group~$\GG$. Moreover, the tau-functions corresponding to the multi-soliton
solutions are expected to involve nilpotent elements of~$\GG$, which is the
origin of their remarkable simple form. 

The tau-function formulation of the Generalized Drinfel'd-Sokolov Hierarchies
of~[\Ref{GEN}] has already been worked out in~[\Ref{TAUTIM}], which, in
fact, has largely inspired our approach. However,
there are two important differences between our results and those
of~[\Ref{TAUTIM}]. Firstly, our approach applies to the affine Toda equations
too, and, secondly, it does not rely upon the use of (level-one) vertex
operator representations. 

At this point, it is worth recalling that the
solutions constructed in section~2 are completely representation-independent. 
In contrast, our definition of tau-functions makes use of a special class of
representations of the Kac-Moody algebra $\gg$ called ``integrable
highest-weight'' representations, which are briefly reviewed in the appendix.
The reason why these representations are called ``integrable'' is the
following. For an infinite-dimensional representation, it is generally
not possible to go from a representation of the algebra $\gg$ to a
representation of the corresponding group $\GG$ via the exponential map
$x\mapsto {\rm e}^x$. However, the construction does work if, for
instance, the formal power series terminates at a certain power of $x$,
or if the representation space admits a basis of eigenvalues of $x$.
These conditions, applied to the Chevalley generators of $\gg$,
single out this special type of representations. 

The generalized tau-functions will be sets of matrix elements of the form
indicated on the right-hand-side of~\solspec, considered as functions of the
group element $h$. They are characterized by the condition that they allow one
to parameterize all the components of $\Theta$, $B$, and $\Upsilon$ required to
specify the solutions~\DPlus\ of the zero-curvature equations~\ZeroCurv. As we
have discussed before, the tau-functions corresponding to the multi-soliton
solutions are expected to have a very simple form. However, in contrast with
the original method of Hirota, we cannot ensure in general that the
equations of the hierarchy become simpler in terms of this new set of
variables.  

First, let us discuss the generalized Hirota tau-functions associated with the
components of $B$. In equation~\solspec, these components can be isolated by
considering the vectors $\ket{\mu_0}$ of an integrable highest-weight
representation $L(\tilde\s)$ of $\ggg$ which are annihilated by all the 
elements in $\gg_{>0}(\s)$, {\it i.e.\/}, $T\ket{\mu_0} = 0$ and 
$\bra{\mu_0}T'=0$ for all $T\in \gg_{>0}(\s)$ and $T'\in \gg_{<0}(\s)$,
respectively. Then, the corresponding tau-functions are defined
as~\note{Since the resulting  relations between tau-functions and components of
the $A_N$'s will be considered as generic changes of variables (see~\kdvtau\
and~\TodaTau), we will not generally indicate the intrinsic dependence of the
tau-functions on the group element $h$.}
$$
\eqalign{
\tau_{\mu_0,\mu_{0}'}(t)\> &=
\> \bra{\mu_{0}'}\Psi^{({\rm vac})}\> h\> {\Psi^{({\rm
vac})}}^{-1}\ket{\mu_{0}} \cr 
&= \> \bra{\mu_{0}'}\> e^{\sum_{N} \varepsilon_N t_N}
\>  h \>  e^{-\sum_{N} \varepsilon_N t_N} \> \ket{\mu_{0}}\>, \cr}
\nfr{TauB}
and, in terms of them, equation~\solspec\ becomes just
$$
\bra{\mu_{0}'}\>  B^{-1}\> \ket{\mu_{0}}\> =\> \tau_{\mu_{0},\mu_{0}'}(t)\>.
\nfr{solspecBF}

By construction, $\gg_0(\s)$ always contains the central element $c$ of the
Kac-Moody algebra, but it is always possible to split the contribution of the
corresponding field in~\solspecBF. Let $s_q\not=0$ and consider the subalgebra
$\overcirc{\ggg}$ of
$\ggg$ generated by the $e_{i}^\pm$ with $i=0,\ldots,r$ but
$i\not=q$, which is a semisimple finite Lie algebra of rank $r$
($\overcirc{\ggg}$ is always simple if $q=0$). Then, $\gg_0(\s) =
\bigl(\gg_0(\s)\cap \overcirc{\ggg}\bigr)
\oplus {\Bbb C}\> c$ and, correspondingly, $B$ can be split as $B= b\> \exp(\nu
\>c)$. Here, $\nu$ is the field along $c$, and $b$ is a function taking values
in the semisimple finite Lie group $\overcirc{G_0}$ whose Lie algebra is
$\gg_0(\s)\cap \overcirc{\ggg}$. Since
$\tilde K=\sum_{i=0}^r k_{i}^\vee \tilde s_i$ is the level of the representation
$L(\tilde\s)$, eq.~\solspecBF\ is equivalent to
$$
\bra{\mu_{0}'}\>  B^{-1}\> \ket{\mu_{0}}\> =\> {\rm
e\/}^{-\nu\> \tilde K}\>\bra{\mu_{0}'}\>  b^{-1}\> \ket{\mu_{0}}\>=\>
\tau_{\mu_{0},\mu_{0}'}(t)\>.
\nfr{CentreA}
Moreover, it is always possible to introduce a tau-function for the field
$\nu$. Let us consider the highest-weight vector $\ket{v_q}$ of the
fundamental representation $L(q)$, which is obviously annihilated by all the
elements in $\overcirc{\ggg}$. Therefore,
$$
\bra{v_{q}}\>  B^{-1}\> \ket{v_q}\> =\> {\rm
e\/}^{-\nu\> k_{q}^\vee}\>= \tau_{v_q,v_q}(t)\equiv \>
\tau_{q}^{(0)}(t)\>,
\nfr{CentreA}
which leads to
$$
\bra{\mu_{0}'}\>  b^{-1}\> \ket{\mu_{0}}\>=\> 
{\tau_{\mu_{0},\mu_{0}'}(t) \over
\Bigl(\tau_{q}^{(0)}(t)\Bigr)^{\tilde K/k_{q}^\vee}}\>\quad  {\rm
and}\quad \nu\>= \> -\> \ln {{\tau_{q}^{(0)}(t)\over k_{q}^\vee}}\>.
\nfr{CentreB}

Finally, recall that the vectors $\ket{\mu_0}$ form a representation of
the semisimple Lie group $\overcirc{G_0}$. Therefore, if $L(\tilde\s)$ is
chosen such that this representation is faithful, eq.~\CentreB\ allows one to
obtain all the components of $b$ in terms of the generalized tau-functions
$\tau_{\mu_{0},\mu_{0}'}$ and $\tau_{q}^{(0)}$. Notice 
that, in this case, the definition of generalized tau-functions coincide 
exactly with the quantities involved in the solitonic specialization of the
Leznov-Saveliev solution proposed in~[\Ref{SOLSPEC}]. 

Let us now discuss the generalized tau-functions associated with the components
of $\Theta$. Consider the gradation $\s$ of $\ggg$ involved in the
definition of the integrable hierarchy. For each $s_i\not=0$, let us consider
the highest-weight vector of the fundamental representation $L(i)$ and define
the (right) tau-function vector
$$
\eqalign{
\ket{\tau_{i}^R(t)}\> & =\> \Psi^{({\rm vac})}\> h\> {\Psi^{({\rm
vac})}}^{-1}\> \ket{v_i} \cr 
& = \>  e^{\sum_{N} \varepsilon_N t_N} \> h \> 
e^{-\sum_{N} \varepsilon_N t_N} \> \ket{v_i}\cr\>.}
\nfr{TauFunc}
Notice that $\ket{\tau_{i}^R(t)}$ is a vector in the representation $L(i)$.
Therefore, it has infinite components, and it will be shown soon that the role
of the Hirota tau-functions will be played by a finite subset of them.
Taking into account that $\ket{v_i}$ is annihilated by all the elements in
$g_{>0}(\s)$, equation~\solspec\ implies
$$
\Theta^{-1}\> B^{-1}\> \ket{v_i}\> =\> \ket{\tau_{i}^R(t)} \>,\quad
i=0,\ldots,r \quad{\rm and}\quad s_i\not=0\>.
\nfr{SolspecT}

The definition~\TauFunc\ is inspired by the tau-function approach
of~[\Ref{KW},\Ref{TAUTIM},\Ref{LUIZ}]. However, in~[\Ref{TAUTIM},\Ref{LUIZ}],
the authors consider a unique tau-function
$\ket{\tau_{\s}(t)} \in L(\s)$. In fact, one could equally consider different
tau-functions $\ket{\tau_{\s'}(t)}$ associated with any integrable
representation $L(\s')$ such that $s_{i}'\not=0$ if, and only if,
$s_i\not=0$. According to  eq.~(A.6),
%\Tensor
all these choices lead to the same results, but ours is the most economical.  

Since, for any integrable representation, the derivation $d_\s$ can be
diagonalized acting on $L(\s)$, these tau-functions vectors can be decomposed as
$$
\ket{\tau_{i}^R (t)}\> =\> \sum_{-j\in {\Bbb Z}\leq0} \ket{\tau_{i}^{R(-j)}
(t)}\>,
\qquad d_i \> \ket{\tau_{i}^{R(-j)} (t)}\> =\> -j \>\ket{\tau_{i}^{R(-j)}
(t)}\>, \nfr{CTau}
where we have used that $\Theta\in \GG_{<0}(\s)$ and $B\in \GG_0(\s)$, and
$d_i$ indicates the derivation corresponding to the gradation with
$s_j=\delta_{j,i}$ (see the appendix). Moreover, the highest-weight vector is an
eigenvector of the subalgebra
$\gg_0(\s)$ and, consequently, of $B$. Therefore,
$$
\ket{\tau_{i}^{R(0)} (t)}\> =\> B^{-1}\> \ket{v_i}\> =\>
\tau_{i}^{(0)} (t) \> \ket{v_i}\>,
\nfr{EienZero}
where, $\tau_{i}^{(0)} (t)$  is a ${\Bbb C}$-function, not
a vector of $L(i)$, whose definition is~\note{To compare with~\solspecBF,
notice that $\ket{\mu_0} = \ket{v_i}$ forms a one-dimensional representation
of $\gg_0({\s})$ and, consequently, $\tau_{v_i,\mu_{0}'}(t)$ vanishes unless
$\ket{\mu_{0}'}=\ket{v_i}$. Therefore, for non-abelian $G_0$, the
required tau-functions $\tau_{\mu_0,\mu_{0}'}(t)$ have to involve the
fundamental integrable representations $L(j)$ corresponding to $s_j=0$, in
contrast with $\ket{\tau_{i}^R(t)}$ (see eq.~\SolspecT).}   
$$
\tau_{i}^{(0)} (t) \>=\> 
\bra{v_i}\> e^{\sum_{N} \varepsilon_N t_N}
\> h\>\>
e^{-\sum_{N} \varepsilon_N t_N}\> \ket{v_i}\>\equiv \> \tau_{v_i,v_i}(t)
\nfr{TauCero}
(compare with eq.~\CentreA). Therefore, eq.~\SolspecT\ becomes
$$
\Theta^{-1}\> \ket{v_i} \> =\> {1\over \tau_{i}^{(0)} (t)}\> \ket{\tau_{i}^R
(t)}\>,
\nfr{TetaTau}
which is the generalization of the eq.~(5.1) of~[\Ref{TAUTIM}] for general
integrable highest-weight representations of $\ggg$. Eq.~\TetaTau\
allows one to express all the components of $\Theta$ in terms of
the components of $\ket{\tau_{i}^R (t)}$ for all $i=0,\ldots,r$ with
$s_i\not=0$ (for instance, by using the positive definite Hermitian form of
$L(i)$). However, it is obvious that only a finite subset of them enter in the
definition of the potentials $A_N$ through eq.~\DPlus.

In exactly the same way, one can introduce another set of ``left''
tau-function vectors through
$$
\bra{\tau_{i}^L(t)}\>  =\> \bra{v_i}\> \Psi^{({\rm vac})}\> h\> {\Psi^{({\rm
vac})}}^{-1}\>,
\efr
which leads to
$$
\bra{v_i}\> \Upsilon \> =\> \bra{\tau_{i}^L (t)}\> {1\over
\tau_{i}^{(0)} (t)}\>,
\efr
and allows one to express all the components of $\Upsilon$ in terms of the
components of $\bra{\tau_{i}^L (t)}$ for all $i=0,\ldots,r$ with
$s_i\not=0$.

Summarising, the generalized Hirota tau-functions of these hierarchies consist
of the subset of functions $\tau_{\mu_0,\mu_{0}'}$ and of components of
$\ket{\tau_{i}^{R}}$ and $\bra{\tau_{i}^{L}}$ required to parameterize all the
components of the potentials $A_N$ in eq.~\DPlus. Then, for the multi-soliton
solutions corresponding to the group element $h$ specified in~\eigenb, their
truncated power series expansion follows from the
possible  nilpotency of the eigenvectors $F_k$ in these representations. For
instance, if
$n=1$ in~\eigenb\ and 
$F_{1}^m\ket{\mu_0} =F_{1}^m\ket{v_i} =0$ unless $m\leq m_1$, then 
$$
\eqalignno{
\tau_{\mu_0,\mu_{0}'}(t)\>& = \> \tilde\tau_{\mu_0,\mu_{0}'}^{0}\>
+\> 
\tilde\tau_{\mu_0,\mu_{0}'}^{1} \>+\> 
\ldots + \tilde\tau_{\mu_0,\mu_{0}'}^{m_1} \cr
& = \sum_{k=0}^{m_1}\> {1\over k!\>}\> {\rm e\>}^{k\> \sum_{N} \> \omega_N
t_N}\>\bra{\mu_{0}'} \>
F_{1}^{k}\> \ket{\mu_0}\>,\quad {\rm and}\cr  
\ket{\tau_{i}^R(t)} \>& = \> \sum_{k=0}^{m_1}\> {1\over k!\>}\> {\rm e\>}^{k\> 
\sum_{N} \> \omega_N t_N}\>F_{1}^{k}\> \ket{v_i}\>. &\nameali{trunca2}\cr}
$$

\chapter{Examples}

The orbits generated by the group of dressing transformations acting on the
vacuum configurations described in section 2 provide solutions of the
generalized mKdV equations of~[\Ref{GEN}], and of the non-abelian affine Toda
equations. In this section we will characterize the appropriated choices for 
$\ss$, and derive the relation between the original variables and their
tau-functions in the simplest cases in order to illustrate the main issues of
our formalism. Moreover, these examples show how the usual definitions
of tau-functions in abelian Toda equations~[\Ref{TIMSOL},\Ref{HIRL}] are
precisely recovered. 

For simplicity, we will restrict ourselves to vacuum solutions associated
with untwisted affine Kac-Moody algebras, although our construction applies
also to the twisted case. Then, it will be convenient to use the realization of
$\gg$ as the central extension of the loop algebra of simple finite Lie
algebra $g$, such that
$$
\eqalignno{
&\gg= \bigl\{ u^{(m)} \>\bigm|\> u\in g\>,\; m \in {\Bbb Z} \bigr\}\>
\oplus
\> {\Bbb C}\> c\>, \qquad \ggg\>\equiv \> g^{(1)} = \gg \oplus {\Bbb C}\>d\>,
\cr
\noalign{\vskip 0.2cm}
&\bigl[ u^{(m)}\> ,\> v^{(n)}\bigr] \>= \> [u\>
,\>v]^{(m+n)}\> +
\> m\>  {\rm Tr\/}(u\> v) \> c\> \delta_{m+n,0}\>, \cr
&\bigl[ d\>, \> u^{(m)}\bigr] \>= \> m\> u^{(m)}\>, \qquad \bigl[ c\>,
\> d\bigr] \> =\> \bigl[ c\>, \> u^{(m)}\bigr]\> =\> 0\>, &
\nameali{Loop} \cr}
$$
where ${\rm Tr\/}(\cdot\; \cdot)$ denotes the Cartan-Killing form of $g$. Then,
the Chevalley generators of $g^{(1)}$ are
$$
e_{i}^\pm \>=\> \cases{E_{\pm \alb_i}^{(0)}\>, & for 
$i=1,\ldots,r\>$,\cr
\noalign{\vskip 0.3cm}
E_{\pm \alb_0}^{(\pm 1)}\>, & for $i=0\>$,\cr}\qquad
h_i \>=\> {2\over \alb_{i}^2}\> \alb_i\cdot \Hb^{(0)} \> +\> c\>
\delta_{i,0}\>,
\nfr{LoopGen}
where $\alb_0 = -\sum_{i=1}^r k_i\> \alb_i$ is minus the highest root
of $g$ normalized as $\alb_{0}^2 =2$, $E_{\alb}$ is the step operator of the
root $\alb$, and
$\Hb$ is an element of the Cartan subalgebra of $g$ ($\Hb$ and $\alb$ live
in the same $r$-dimensional vector space).

We will also use the notation
$$
\gg_{\leq k}(\s)= \bigoplus_{i\leq k} \gg_i(\s)\>, \qquad
\gg_{\geq k}(\s)= \bigoplus_{i\geq k} \gg_i(\s)\>,
\efr
and denote by ${\rm P}_{\geq k[\s]}$ and ${\rm P}_{<k[\s]}$ the projectors 
onto $\gg_{\geq k} (\s)$ and $\gg_{<k}(\s)$, respectively.

Different choices of the subset $\ss$ introduced in~\Commute\ lead to solutions
of different integrable hierarchies. However, particularly interesting
vacuum solutions arise when $\ss$ is a subset of a Heisenberg subalgebra of
$\ggg$, which, up to the central element $c$, are special types of
maximally commuting subalgebras whose precise definition can be found
in~[\Ref{KP}]. In particular, when $\ggg=g^{(1)}$, they
correspond to the affinization of a Cartan subalgebra of $g$ by means of an
inner automorphism. This implies that the inequivalent Heisenberg subalgebras
of $g^{(1)}$ are classified by the conjugacy classes of the Weyl group of
$g$~[\Ref{KP}]. Consequently, their structure is 
$$
\Heis = {\Bbb C}\> c + \sum_{i\in I_{[w]} + {\Bbb Z}N_{[w]}} {\Bbb
C}\Lambda_i\>, \qquad [\Lambda_i, \Lambda_j]=i\> c\> \delta_{i+j,0}\>,
\nfr{Heisenberg}
where $[w]$ denotes a conjugacy class of the Weyl group of $g$, and
$I_{[w]}$ is a set of $r$ integers $\geq0$ and $<N_{[w]}$. The set
$\{c,\> \Lambda_i\bigm| i\in I_{[w]} + {\Bbb Z}N_{[w]}\}$ is a basis of
$\Heis$ whose elements are graded with respect to the associated
$[w]$-dependent gradation $\sw=(s^{w}_0,\ldots,s^{w}_r)$. The gradation
$\sw$ fixes the set $I_{[w]}$ and the integer $N_{[w]} = \sum_{i=0}^{r}
k_i s_{i}^{w}$, where $k_0=1$, $k_1, \ldots, k_r$ are the labels of
the extended Dynkin diagram of $g$, which also specify its highest root
$\alb_{0}$. 

\section{Generalized mKdV and KdV hierarchies.}

Let $\Lambda_i$ be an element of a Heisenberg subalgebra $\Heis$ of $g^{(1)}$
whose $\s^w$-grade is $i>0$, and consider the subalgebra
$$
\ss= {\rm Cent\;} \bigl(\Ker ( \ad \Lambda_i)\bigr) \cap
\gg_{\geq0} (\sw) \subseteq \Heis \cap \gg_{\geq0} (\sw)\>,
\nfr{VacmKdV}
where by ${\rm Cent\;}(\bullet)$ we mean the subalgebra of $(\bullet)$
generated by the elements which commute, up to the central element $c$, with
all elements of $(\bullet)$. Then,~$\ss$ gives rise to the vacuum solution
$$
A_N^{({\rm vac})}\> =\> \Lambda_N \>+ \> \rho_N(t)\> c\>,
\nfr{VacKdV}
which is labelled by the set of integers $N$ such that $\ss \cap
\gg_N(\sw)$ is not empty, and where $\Lambda_N \in \ss \cap \gg_N(\sw)$. To
compare with eq.~\vacpot, $N_-=0$, $N_+=N$, and $c_{N}^j =\delta_{j,N}$.
According with eq.~\DPlus, the orbit of solutions generated by the
group of dressing transformations acting on this vacuum consists of the Lax
operators
$$
\eqalign{
{\cal L}_{N}^h \>&=\> \Theta \bigl({\partial\over \partial t_N}\> -\> 
\Lambda_N \bigr)\Theta^{-1}\>- \> \rho_N(t)\> c \cr
&=\> {\partial\over \partial t_N}\> -\> \Pos\bigl(\Theta\>
\Lambda_N\>\Theta^{-1}\bigr) \>- \> \rho_N(t)\> c\>, \cr}
\nfr{KdVOrbit}
where $\Theta$ is defined in~\Teta\ and it will be understood as a function
of the group element~$h$. 
 
Then, the eqs.~\KdVOrbit\ provide solutions
for one of the generalized Drinfel'd-Sokolov hierarchies of~[\Ref{GEN}] (see
also~[\Ref{TAUTIM}]). In particular, for the generalized mKdV hierarchy
associated with the Lax operator
$$
L = {\partial \over\partial x} - \Lambda - \widetilde{q}  \equiv {\cal L}_i\>,
\nfr{LaxDS}
where $\Lambda= \Lambda_i$, $x=t_{i}$, and
$$
\widetilde{q} \in \gg_{\geq0}(\sw) \cap \gg_{<i}(\sw)\>.
\nfr{LaxViejo}
In order to prove it, let us briefly summarize its construction. It is based
on the existence of a unique transformation such that
$$
\Phi\> L\> \Phi^{-1} \>= \> {\partial \over\partial x} - \Lambda -
h\>,
\nfr{Abel}
where $\Phi =\exp y$ with $y\in \Im (
\ad \Lambda) \cap \gg_{<0}(\sw)$, and $h\in \Ker (\ad \Lambda) \cap
\gg_{<i}(\sw)$ are local functionals of the components of $\widetilde{q}$ and
their $x$-derivatives (notice that $\Ker (\ad \Lambda)$ is non-abelian in
general). There is a difference between the situation in~[\Ref{GEN}] and the
situation in~[\Ref{TAUTIM}] and here. Here, $\widetilde{q}$ may have
a component along the central element of $g^{(1)}$, say $\widetilde{q} = q +
q_c\> c$, where $q$ is the component of $\widetilde{q}$ in the loop algebra of
$g$. Then, it is straightforward to show that $\Phi$ and $h - q_c\> c$ depend
only on the components of $q$. The hierarchy consists of an infinite set of
commuting flows associated with the elements $\Lambda_N$ in $\ss$, and they are
defined by the zero-curvature equations
$$
\bigl[{\partial \over \partial t_N}\>- \> \Pos(\Phi^{-1}\> \Lambda_N
\>\Phi)\>, \> L\bigr]\>=\>0\>.
\nfr{FlowmKdV}
Moreover, since $\Phi$ is a differential polynomial of $q$ and it does not
depend on $q_c$, this flow equation induces a flow equation for $q$ which can be
written in the form
$$
{\partial q\over \partial t_j} = F_j \bigl( q, {\partial q\over
\partial x}, {\partial^2 q\over
\partial x^2}, \ldots\bigr)\>,
\efr
for some polynomial functions $F_j$. In contrast, the corresponding equation
for the component along the central element of $g^{(1)}$ is $\partial_N q_c =
-\partial_x  (\Phi^{-1}\> \Lambda_N \>\Phi)_c$, where $(\bullet)_c$ is the
component of
$\bullet$ along ${\Bbb C}\> c$. Since $\Phi$ depends only on $q$, this shows
that
$q_c$ is not a real degree of freedom, which is the reason why these
hierarchies can be  associated both with Kac-Moody or loop algebras.

Let us consider the mKdV hierarchy~\FlowmKdV\ constrained by the condition
that $h\in \Ker (\ad \Lambda) \cap
\gg_{<0}(\sw)$. Since all the vanishing components of $h$ in $\gg_{\geq0}
(\sw)$ are functionals of $\widetilde{q}$, this implies a constraint on the mKdV
field, and it is easy to check that it is compatible with the flow equations.
Then, using~\Abel\ and introducing a non-local functional $\chi$ of $q$ such
that $\chi\in \exp\bigl(\Ker (\ad \Lambda) \cap
\gg_{<0}(\sw)\bigr)$ and $\partial_x \chi \>{\chi}^{-1} =h$, the
Lax operator becomes
$$
\eqalign{
L \>& = \Theta\bigl({\partial \over\partial x} - \Lambda\bigr) \Theta^{-1}\>
+\> ({\chi}\Lambda {\chi}^{-1})_c \cr
& = {\partial \over\partial x} - \Pos\bigl( \Theta\Lambda
\Theta^{-1}\bigr)\> +\> (\chi\Lambda \chi^{-1})_c \>, \cr}
\nfr{DSNuevo}
where $\Theta = \Phi^{-1}\> \chi \in \GG_-(\sw)$. Moreover, the Lax operators
that define the flows of the hierarchy can be written as
$$
{\partial \over \partial t_N}\>- \> \Pos(\Phi^{-1}\> \Lambda_N
\>\Phi)\> =\>  {\partial \over \partial t_N}\>- \> \Pos(\Theta\> \Lambda_N
\>\Theta^{-1})\> -\> (\chi^{-1}\> \Lambda_N
\>\chi)_c\>.
\efr
This, compared with~\KdVOrbit, shows that the orbit generated by the group of
dressing transformations acting on the vacuum solution~\VacKdV\
actually consists of solutions of the generalized mKdV hierarchy associated 
with the Lax operator~\LaxDS.

Eqs.~\LaxDS\ and \DSNuevo\ provide the change of variables between
$\widetilde{q}$ and the components of $\Theta$, and they show that only the
finite number of terms of
$\Theta$ with $\sw$-grade ranging from $-i$ to $-1$ are required. Therefore,
in this case, the generalized tau-functions correspond to a finite set of
components of the vectors $\ket{\tau_{i}^R(t)}$ in the fundamental integrable
representations $L(i)$ such that $s^{w}_i\not=0$. Let us remark that this case
is covered by the results of~[\Ref{TAUTIM}] only if these representations are of
level-one, which means that that $s^{w}_i\not=0$ only if $k_{i}^\vee=1$.

As an specific example, let us discuss the Drinfel'd-Sokolov generalized mKdV
hierarchies associated to a simple finite Lie algebra $g$. They are recovered
from the principal Heisenberg subalgebra, which is graded with respect to the
principal gradation $\s_{\rm p} =(1,1,\ldots,1)$, and the Lax operator \LaxDS\
where, in this case,
$$
\Lambda \>=  \> \Lambda_1 \>= \> \sum_{i=0}^r e_{i}^+ \quad {\rm and} \quad
\widetilde{q}= \sum_{i=0}^r q_i\> h_i\> .
\nfr{mKdVg}

The change of variables between $\widetilde{q}$ and the components of $\Theta$
follows from~\LaxDS\ and \DSNuevo\ by writing 
$$
\Theta = \exp\left( \theta_{-1} + \cdots\right)
\in \GG_-(\s_p)\>,
$$ 
with $\theta_{-1}\in \gg_{-1}(\s_p)$, which leads to
$$
\widetilde{q}\> =\> [\theta_{-1}\> ,\> \Lambda]\>.
\nfr{Change}

The relation between $\theta_{-1}$ and the tau-functions has to be
obtained from eq.~\TetaTau. Consider the decomposition $\theta_{-1} =
\sum_{i=0}^{r} a_i e_{i}^{-}$ for some functions
$a_j$, then~\TetaTau\ implies that
$$
\sum_{i=0}^r a_i\> e_{i}^{-}\> \ket{v_j}\> = \>-\>{1\over
\tau_{j}^{(0)}(t)} \> \ket{\tau_{j}^{R(-1)}(t)}\>.
\nfr{MinOne}
Moreover, since the times are labelled by positive
integers (the exponents of $g$ plus its Coxeter number times some non-negative
integer) and $\Lambda_N$ annihilates the highest-weight vectors for
$N>0$, eq.~\TauFunc\ reduces to
$$
\ket{\tau_{j}^R(t)}\> =\> {\rm e\/}^{\sum_{N} t_{N}\> \Lambda_N} \>   
h\> \ket{v_j}\>,
\nfr{TauJ}
and eq.~\TauCero\ becomes
$$
\tau_{j}^{(0)}(t) \> =\> \bra{v_j}\> {\rm e\/}^{\sum_{N} t_{N}\> \Lambda_N} \>
h\> \ket{v_j}\>.
\nfr{TauCeroDS}
Then, eqs.~\MinOne\ and \TauJ\ imply that
$$
a_j\> = \> -\>{1\over
\tau_{j}^{(0)}(t)} \> \bra{v_j}\>e_{j}^+\> {\rm e\/}^{\sum_{N} t_{N}\>
\Lambda_N} \> h\> \ket{v_j}\>,
\efr
but $\bra{v_j}\> e_{j}^+ \>= \>\bra{v_j} \Lambda_1$, which, taking
into account~\TauCeroDS, finally leads to
$$
a_j \>= \>-\> {\partial\over \partial x}\> \ln\>
\tau_{j}^{(0)}(t)\>.
\nfr{mKdVFin}

Therefore, using~\Change, the resulting change of variables is
$$
\eqalign{
\widetilde{q} \>&=\>  \sum_{i=0}^r\> {\partial\over \partial x}\> \ln\>
\tau_{i}^{(0)}(t) \> h_i\cr
&= \>  \sum_{i=1}^r \> {\partial\over \partial x}\> \ln\biggl(
{\tau_{i}^{(0)}(t)\over
\tau_{0}^{(0)\>k_{i}^{\vee}}(t)}\biggr)\> {2\over \alb_{i}^2}\>
\alb_i \cdot \Hb^{(0)}\> +\>
\tau_{0}^{(0)}(t)\> c \>,\cr}
\nfr{mKdVTot}
where $k_{i}^\vee = (\alb_{i}^2/2) k_i$, which shows that
$\tau_{0}^{(0)}(t), \ldots, \tau_{r}^{(0)}(t)$ are the
Hirota tau-functions in this case.

In general, when $\ss\subset \gg_{\geq0}(\s)$, as in~\VacmKdV, the construction
presented in sections~2 and~3 can still be generalized by introducing an
auxiliary gradation $\s^\ast \preceq\s$ with respect to the partial ordering
of~[\Ref{GEN}]. Then, the new Lax operators would be defined such that 
$$
A_N \in \gg_{\geq0}(\s^\ast) \cap \gg_{\leq N}(\s)\>,
\nfr{newpot}
and the analogue of the dressing transformation~\Dressing\ involves the
factorization
$$
\Psi\> h\> \Psi^{-1}\> \in \GG_{-}(\s^\ast)\> \GG_0(\s^\ast)\>
\GG_{+}(\s^\ast)\>.
\nfr{newdecomp}
Now, if $\ss$ is of the form given by eq.~\VacmKdV, the orbit of the
vacuum solution~\VacKdV\ provides solutions for the generalized
(partially modified) KdV hierarchies of~[\Ref{GEN}], and it leads to the
corresponding generalizations of the Hirota tau-functions~[\Ref{TAUTIM}]. 

In addition, the form of the new Lax operator~\newpot, is invariant under the 
gauge transformations~[\Ref{GEN}]
$$
\cl_N  \> \mapsto\> U \cl_N U^{-1} = U \biggl( {\pa \, \over \pa t_N} - A_N 
\biggr) U^{-1}
\nfr{GaugeT}
where $U$ is an exponentiation of elements of the algebra
$$
P \equiv \gg_{0}(\s^\ast) \cap \gg_{\leq 0}(\s)\>
\efr
In terms of the associated linear problem, the gauge transformation~\GaugeT\
corresponds to $\Psi\mapsto U\Psi$, with $U\in P$. This opens the
possibility of considering different gauge equivalent definitions of the 
dressing
transformation~\Dressing, (with the decomposition~\Factor\  being replaced
by~\newdecomp), \eg,
$$
\eqalign{
\Psi\> \mapsto\> U\Psi^h & =[(\Psi\> h\> \Psi^{-1})_- U^{-1}]^{-1}\> 
\Psi\cr
& = [U (\Psi\> h\>
\Psi^{-1})_0]\> (\Psi\> h\> \Psi^{-1})_+ \>\Psi\> h^{-1}\>,\cr}
\nfr{DressGauge}
leading to gauge equivalent solutions of the zero-curvature equations. It is
worth pointing out that the group of gauge tranformations~\GaugeT\ is not
trivial even if $\s^\ast =\s$. However, along the paper, we only consider 
the dressing transformations defined by~\Dressing, which is equivalent to a
gauge fixing prescription for the transformations~\GaugeT\ (an alternative
prescrition is used, \eg, in~[\Ref{BABT}]).

\section{Generalized non-abelian Toda equations.}

Now, for a given Heisenberg subalgebra $\Heis$ of $g^{(1)}$, let us choose a 
positive integer~$l$ and consider the vacuum solution
$$
A_{l}^{({\rm vac})}\> =\> \Lambda_l\>, \quad
A_{-l}^{({\rm vac})}\> =\> \Lambda_{-l}\> +\> l\> t_l\> c\>,
\nfr{todavacpot} 
associated to the subalgebra generated by $ \Lambda_{\pm l} \in \Heis \cap
\gg_{\pm l}(\sw)$. In~\vacpot, this solution corresponds to $l_+ =l$, $l_-=0$,
$(-l)_+=0$,
$(-l)_-=-l$, $c_{l}^j= \delta_{j,l}$, $c_{-l}^j= \delta_{j,-l}$, $\rho_l=0$,
and $\rho_{-l}=l\>t_l$, and it is equivalent to the following solution of the
associated linear problem
$$
\Psi^{({\rm vac})}\> =\> \exp\Bigl( \Lambda_l\> t_l \>+\> \Lambda_{-l}\> t_{-l}
\>+\> {1\over2}\> l\> t_l\> t_{-l}\> c\Bigr)\>.
\efr
   
If $l>1$, we will only be interested in the orbit of solutions generated by the
dressing transformations associated with the elements of the subgroup
$\GG^{(l)}$ formed by exponentiating the subalgebra
$$
\gg^{(l)} = \bigoplus_{k\in {\Bbb Z}} \gg_{kl}(\sw)\>.
\nfr{TodaSub}
However, let us indicate that the orbit generated by the full Kac-Moody group
acting on~\todavacpot\ provides solutions for the generalized affine 
non-abelian
Toda equations of~[\Ref{SAVGERV}]. Then, $\Psi^{({\rm vac})}$, $h$, and,
consequently, $\Psi^{({\rm vac})}\> h\> {\Psi^{({\rm vac})}}^{-1}$ are in
$\GG^{(l)}$, which implies
$$
\eqalignno{
&\Theta \>=\> \exp \biggl( \sum_{k\in {\Bbb Z}>0} \theta_{-kl}\biggr)\>=\> 1
+\theta_{-l} + \cdots \>\in
\GG^{(l)}_-(\sw)\>, \quad \theta_{-kl} \in \gg_{-kl}(\sw)\>,\cr
\noalign{\vskip 0.3cm}
&\Upsilon = \exp \biggl(\sum_{k\in {\Bbb Z}>0} \zeta_{\>kl}\biggr)\>=\> 1
+\zeta_{l} + \cdots \>\in
\GG^{(l)}_+(\sw)\>, \quad \zeta_{\>kl} \in \gg_{kl}(\sw)\>. &
\nameali{Compon}\cr} 
$$
Therefore, using eqs.~\DPlus, the orbit of solutions generated by the
group of dressing transformations acting on~\todavacpot\ is given by 
$$
\eqalignno{
A_{l}^h &\> =\> \Lambda_l \>+\> [\theta_{-l}, \Lambda_l] &
\nameali{Tplusone} \cr
& \>=\> \Lambda_l \>-\> B^{-1} \partial_{l} B \>, &
\nameali{Tplustwo}\cr
\noalign{\vskip 0.3cm}
A_{-l}^h &\>=\> \Lambda_{-l} \>+\> \partial_{-l} \theta_{-l} \>+\> 
l\> t_{l}\> c \> & \nameali{Tminone} \cr
&\>=\> B^{-1}\Lambda_{-l} B \>+\> l\> t_{l}\> \>c \>.& \nameali{Tmintwo}
\cr}
$$

Using eqs.~\Tplustwo\ and~\Tmintwo, the zero-curvature equation $[{\cal
L}_{l}^h, {\cal L}_{-l}^h]=0$ becomes
$$
\partial_{l}^-\bigl( B^{-1} \partial_{l}^+ B \bigr) \>= \>
\bigl[ \Lambda_l\> ,\> B^{-1}\Lambda_{-l} B \bigr] - l\> c\>,
\nfr{TodaEq}
which shows that $\widehat{B} = e^{l \> t_{l}\> t_{-l}\>  c}\> B$ is a
solution of a generalized non-abelian affine Toda equation where
$\pm t_{\pm l}$ play the role of the light-cone variables $x_\pm= x\pm t$, and
$\widehat{B}$ is the Toda
field~[\Ref{LS},\Ref{OLTUR},\Ref{UNDER},\Ref{LUIZ},\Ref{CAT}]. It is well
known that these equations can be understood as the classical
equations-of-motion of certain two-dimensional relativistic field theories, and
Lorentz invariance is manifested through the symmetry transformation $x_\pm
\rightarrow \lambda^{\pm1} x_\pm$. 

However, the solutions provided by the dressing transformation method satify
additional constraints. Actually, the  comparison of~\Tplusone\ with \Tplustwo\
shows that
$$
B^{-1}\> \partial_l\> B \> =\> [\Lambda_l\>, \> \theta_{-l}]\in \Im(\ad
\Lambda_l)\>,
\nfr{GaugeA}
and~\DPlus\ and~\Tmintwo\ implies that
$$
\partial_{-l}\> B \>B^{-1}\>  =\> -\> [\Lambda_{-l}\>, \> \zeta_{-l}]\in
\Im(\ad \Lambda_{-l})\>.
\nfr{GaugeB}

These constraints break the well known chiral symmetry 
$$
B\mapsto h_-(t_{-l})\> B\> h_+(t_{l})
\nfr{ConstTod}
of the affine Toda equation~\TodaEq, where $h_\pm(t_{\pm l})$ take values in
the subgroups $\GG_{\pm}$ formed by exponentiating the subalgebras $\Ker(\ad
\Lambda_{\pm l})\cap \gg_0(\sw)$. Moreover, eqs.~\GaugeA\ and~\GaugeB\ have a
nice interpretation as gauge-fixing conditions for certain local (chiral)
symmetries of the underlying two-dimensional field
theory~[\Ref{LUIZ},\Ref{PARK}]. 

The dressing transformation method allows one to relate the resulting
solutions of the non-abelian Toda~\TodaEq\ and generalized mKdV~\FlowmKdV\
equations. The crucial observation is that the Lax operator ${\cal L}_{l}^h$
corresponding to~\Tplustwo\ can also be viewed as the Lax operator of a
generalized mKdV hierarchy. Then, according to~\LaxDS, the relation is
$\Lambda=\Lambda_l$, $x=t_l \equiv x_+$, and $\widetilde{q} = -B^{-1}
\partial_l B$. This provides a (locally) non-invertible map from solutions of 
the
non-abelian Toda equation~($B$) into solutions of the mKdV
equation~($\widetilde{q}$), which generalizes the known relation between
solutions of the sine-Gordon and mKdV equations. However, if $l>1$, notice that
the resulting mKdV Lax operator is constrained by the condition $\widetilde{q}=
-B^{-1} \partial_l B\in \gg_{0}(\sw)$, a constraint that is compatible only
with a subset of the flows that define the mKdV hierarchy. In order to make
this relation concrete, let us consider the subalgebra
$$
\ss^\dagger\> =\> \Bigl[{\rm Cent\/}\bigl(\Ker(\ad \Lambda_l)\bigr) \cap
\bigoplus_{k\in {\Bbb Z}\geq0} \gg_{kl}(\sw)\Bigr] \> \cup\> {\Bbb C}
\> \Lambda_{-l}\>,
\nfr{Mix}
and the associated vacuum solution 
$$
A_{kl}^{({\rm vac})}\> =\cases{\Lambda_{kl} &if $k\geq0\>$, \cr 
\Lambda_{-l}\> +\> l\> t_l\> c &if $k=-1\>$.\cr}
\nfr{TandKdV}
Then, the orbit generated by the group of dressing transformations induced by
the elements of $\GG^{(l)}$ provide joint solutions of both the non-abelian 
Toda
equation~\TodaEq\ and the mKdV hierarchy of equations restricted to the
flows generated by the times
$t_{kl}$.~\note{A similar construction can be
used to connect the solutions of the full generalized mkdV hierarchies with the
solutions of the generalized non-abelian Toda equations of~[\Ref{SAVGERV}].}
Recall that, both in the mKdV and non-abelian Toda equations, the solutions
provided by the dressing transformation method satisfy additional constraints.
In the generalized mKdV equations, these constraints are given by the condition
that $h\in \gg_{<0}(\sw)$ (see eq.~\Abel\ and the discussion leading
to~\DSNuevo). As for the non-abelian Toda equation, the so obtained solutions
satisfy the identities~\GaugeA\ and~\GaugeB. However, since, in this case,
$\widetilde{q} = -B^{-1} \partial_l B\in \gg_{0}(\sw)$,  eq.~\Abel\ implies
that $h\in \gg_{\leq0}(\sw)$, and that
$ \widetilde{q} = {\rm P}_{0[\sw]}(h) +[\Lambda_l, \varphi_{-l}]$, where 
$\Phi = 1+  \varphi_{-l} +\cdots$. All this shows that the constraint $h\in
\gg_{<0}(\sw)$ corresponds precisely to eq.~\GaugeA. In contrast, eq.~\GaugeB\
involves the time~$t_{-l}$ and, hence, it does not have an interpretation in
terms of the mKdV hierarchy associated with $\widetilde{q}$.

Finally, let us discuss the tau-functions for these non-abelian
Toda hierarchies. Eqs.~\Tplustwo\ and~\Tmintwo\ show that the non-abelian Toda
equation~\TodaEq\ is a partial differential equation for the Toda field $B$.
Actually, this remains true if we consider the hierarchy of equations
associated with the vacuum solution~\TandKdV. Therefore, in this case, the
generalized Hirota tau-functions correspond to the set of variables
$\tau_{\mu_0,\mu_{0}'}(t)$, and the change of variables is provided by
eq.~\solspecBF. This shows that the vectors $\ket{\mu_0}$ have to be chosen in
some integrable representation of
$g^{(1)}$ such that they form a faithful representation of $\GG_0(\sw)$.
It is worth mentioning that eqs.~\Tplusone\ and~\Tminone\ suggest the 
possibility
of describing these hierarchies in terms of the components of $\Theta$.
Actually, those equations manifest the existing relations between the
tau-functions $\tau_{\mu_0,\mu_{0}'}(t)$ and the components of
$\ket{\tau_{i}^R(t)}$ that provide the tau-functions of the associated mKdV 
hierarchy. However, these relations are non-local and, therefore, not very
useful in practice in the general case.

Eq.~\TauB\ shows that the proposed generalized tau-functions
$\tau_{\mu_0,\mu_{0}'}(t)$ are precisely the matrix elements involved in the
solitonic specialization of the Leznov-Saveliev solution
of~[\Ref{SOLSPEC}]. Therefore, using the map between the solutions of
non-abelian Toda and mKdV equations, the dressing transformation method relates
the resulting orbit of solutions of the latter with the solitonic specialization
of the Leznov-Saveliev solution, originally formulated in the context of affine
Toda equations. Then, since it can be justified that the solitonic
specialization singles the soliton solutions out from the general
Leznov-Saveliev solution~[\Ref{TUROLA},\Ref{TUROLB},\Ref{DOV}], the observed 
relation supports
the conjecture that the orbits of solutions generated by the group of
dressing transformations actually contain all the multi-soliton solutions of the
equations.

The simplest example is provided by the abelian affine Toda equations, which
are related with the Drinfel'd-Sokolov mKdV hierarchies discussed in
section~4.1. Therefore, they are recovered from the principal
Heisenberg subalgebra and $l=1$, with
$$
\Lambda_1\> =\> \sum_{i=0}^r \> e_{i}^+ \quad {\rm and} \quad
\Lambda_{-1}\> =\> \sum_{i=0}^r\> k_{i}^\vee\> e_{i}^-\>.
\efr
The Toda field takes values in $\gg_0(\s_p)$, which is generated by $h_0,
\ldots, h_r$. This implies that   
$$
B\> =\> \exp\left(- \sum_{i=0}^r \phi_i\> h_i\right)\> =\>
\exp\left(- \> \pp\cdot \Hb^{(0)}\> - \> \phi_0\> c\right)\>,
\efr
and, hence, the generalized Hirota equations are just $\tau_{v_i, v_i}(t) =
\tau_{i}^{(0)}(t)$, for $i=0, \ldots, r$. The relation between the components of
$B$ and the tau-functions follows from eq.~\solspecBF\
$$
\eqalign{
\bra{v_i} B^{-1} \ket{v_i} \> = & \> {\rm e\/}^{\phi_i}\cr
 = & \> \tau_{i}^{(0)}(t)\>, \quad {\rm for} \quad i=0,\ldots, r\>, \cr}
\efr
which leads to
$$
\pp\> =\> \sum_{i=0}^r \> \ln\tau_{i}^{(0)}(t)\> {2\over
\alb_{i}^2}\> \alb_i\>, \quad \phi_0 \> =\>\ln\>
\tau_{0}^{(0)}(t)\>,
\efr
which, compared with~\mKdVTot, exhibits the relation with the mKdV
hierarchies, and agrees with the change of variables~\TodaTau\ used
in~[\Ref{TIMSOL},\Ref{HIRL},\Ref{mm},\Ref{MAX}].

\chapter{Conclusions}

In this paper, we have studied a special type of solutions of a large class of
non-linear integrable zero-curvature equations. The class of integrable models
is constructed from affine (both twisted and non-twisted) Kac-Moody algebras,
and is characterized by exhibiting trivial solutions, referred to as ``vacuum
solutions'', such that the corresponding Lax operators take values in some
abelian subalgebra up to the central term. Then, we have considered the orbits
of solutions generated by the group of dressing transformations acting
on those vacua. It is important to remark that the relevant integrable models
are not constructed explicitly. In contrast, their zero-curvature equations are
found only as the equations satisfied by this particular class of
solutions. This is similar to the tau-function approach of~[\Ref{KW}] where the
equations defining the integrable hierarchy of bi-linear Hirota equations are
derived from the property that their solutions lie in the orbit of a
highest-weight vector generated by a Kac-Moody group. The resulting class of
integrable models include the generalizations of the Drinfel'd-Sokolov
hierarchies of mKdV (and KdV) type constructed in~[\Ref{GEN}], and the
generalizations of the sine-Gordon equation known as abelian and non-abelian
affine Toda equations~[\Ref{LS},\Ref{LUIZ},\Ref{SAVGERV},\Ref{UNDER}].

The motivation for studying this particular type of solutions is to find the
generalizations of the Hirota tau-functions for the relevant integrable
systems. First of all, it is generally assumed that the orbit of solutions
generated by the group of dressing transformations contain all the
multi-soliton solutions of the integrable hierarchy~[\Ref{BABT}]. Then,
according to the method of Hirota, the generalized tau-functions provide an
alternative set of variables that largely simplify the task of constructing
the multi-soliton solutions. In this case, we have identified those new
variables with specific matrix elements evaluated in the integrable
highest-weight representations of the Kac-Moody algebra. 

In particular, for the generalizations of the Drinfel'd-Sokolov
hierarchies of mKdV (and KdV) type, our results constitute a generalization
of the results of~[\Ref{TAUTIM}] to the general case when the relevant
integrable representations are neither of level-one nor of vertex type.
Moreover, for the non-abelian affine Toda equations, our results show that
the suitable generalizations of the Hirota tau-functions correspond to the
matrix elements involved in the solitonic specialization of the
general Leznov-Saveliev solution~[\Ref{SOLSPEC}]. Actually, this is a
remarkable result since it links the orbits of solutions under consideration
with the solitonic specialization of~[\Ref{SOLSPEC}]. Then, since the
solitonic specialization arises as a prescription to single the multi-soliton
configurations out from the general solution, our result supports the
conjecture that all the multi-soliton solutions lie in the orbit generated by
the group of dressing transformation acting on some vacuum.

In this paper we have only considered integrable systems of zero curvature
equations constructed from Kac-Moody algebras. However, there are many other
important integrable hierarchies formulated by means of pseudo-differential
operators, and it would be interesting to investigate the implications of our
results for the definition of generalized Hirota tau-functions in those
cases~[\Ref{DIC}]. In particular, it has been recently shown in
ref.~[\Ref{FMDOS}] how matrix generalizations of both the Gelfand-Dickey and
the constrained KP hierarchies can be recovered from the construction
of~[\Ref{GEN}]. Therefore, at least in these important cases, it should be
possible to translate directly our definition of tau-functions into the context
of those integrable systems.
  
\bjump\bjump

\centerline{{\bf Acknowledgements}}

One of us (L.A.F.) is very grateful to the hospitality received at the 
Departamento de F\'\i sica de Part\'\i culas of the Universidad de Santiago de 
Compostela, Spain and the International Centre for Theoretical Physics - ICTP, 
Trieste, Italy, where part of this work was done. L.A.F. is grateful to
H.~Aratyn, H. Blas Achic, J.F. Gomes and A.H. Zimerman for many helpful
discussions. J.L.M. would like to thank Prof. D.I.~Olive for clarifying the
interpretation of the solitonic specialization, T.J.~Hollowood for his early
collaboration in this work, and M.A.C.~Kneipp for useful conversations. J.L.M.
and J.S.G. are supported partially by CICYT (AEN96-1673) and DGICYT (PB93-0344).

\bjump

\appendix{Appendix: Integrable Highest-Weight Representations}

An affine Kac-Moody algebra~$\ggg$ of rank $r$ is defined by a generalized
Cartan matrix $a$ of affine type of order $r+1$ (and rank~$r$), and is
generated by $\{h_i, e_{i}^\pm, i=0, \ldots, r\}$ and $d$ subject to the
relations~[\Ref{KBOOK}]
$$
\eqalign{
[h_i\> ,\> h_j]\> =0\>, \quad & [h_i\> ,\>e_{j}^\pm] \> =\> \pm a_{i,j}\>
e_{j}^\pm\>, \cr
[e_{i}^+\>, \> e_{j}^-] \> =\> \delta_{i,j}\> h_i \>, \quad & (\ad
e_{i}^\pm)^{1-a_{i,j}} \> (e_{j}^\pm)\> =\> 0\>, \cr
[d\>, \> h_i]\> =0\>, \quad & [d\>, \> e_{i}^\pm]\> =\> \pm
\delta_{i,0}\> e_{0}^\pm\>. \cr}
\efr
The elements $e_{i}^\pm$ are Chevalley generators, and $\{h_0,
\ldots, h_r, d\}$ span the Cartan subalgebra of $\ggg$. The algebra
$\ggg$ has a centre ${\Bbb C}\> c$ generated by the central element
$c=\sum_{i=0}^{r} k_{i}^\vee \> h_i$ where $k_{i}^\vee$ are the labels of the
dual Dynkin diagram of $\ggg$ (the dual Kac labels), and in all cases
$k_{0}^\vee =1$.

The different $\Bbb Z$-gradations of~$\ggg$ are labelled by sets $\s=(s_0,
\ldots, s_r)$ of non-negative integers. Then the gradation is induced by a
derivation $d_\s$ such that
$$
[d_\s \>, \> h_i]\> = \> [d_\s \>, \> d] \>= \> 0\>, \quad
[d_\s \>, \> e_{i}^\pm] \> =\> \pm s_i e_{i}^\pm\>.
\efr
In particular, the derivation $d$ corresponds to the, so called, homogeneous
gradation
$\s=(1,0,\ldots,0)$. 
 
The definition of integrable representations makes use of the
following property. An element $x\in \ggg$ is said to be ``locally
nilpotent'' on a given representation if for any vector $\ket{v}$ there
exists a positive integer $N_v$ such that $x^{N_v} \ket{v}=0$. Then,
an integrable highest-weight representation $L(\s)$ of $\ggg$ is a
highest-weight representation of $\ggg$ where the Chevalley generators are
locally nilpotent~[\Ref{KBOOK}]. It can be proven that
$L(\s)$ is irreducible and that $\ket{v_\s}$ is the unique highest-weight vector
of $L(\s)$. 

The highest-weight vector of $L(\s)$ can be labelled by a gradation,
$\s=(s_0,s_1,\ldots,s_r)$ such that
$$ 
\eqalignno{
& e_{i}^+\> \ket{v_\s} \>= \> \left(e_{i}^-\right)^{s_i +1} \> \ket{v_\s}
\>=\>0 \>,  &\nameali{Nilp}\cr 
& h_i \> \ket{v_\s} \>= \> s_i \> \ket{v_\s} \>, \qquad d_\s \>
\ket{v_\s} =0\>,& \nameali{Cartan}\cr}
$$
for all $i=0,\ldots,r$. Notice that the eigenvalue of $d_\s$ is
arbitrary, and that $d_\s$ can be diagonalized acting on $L(\s)$. The
eigenvalue of the centre $c$ on the representation $L(\s)$ is known as
the level $k$ 
$$
c\> \ket{v_\s} \>= \> \sum_{i=0}^r k_{i}^{\vee}\> h_i \> \ket{v_\s} \>= 
\> \left(\sum_{i=0}^r k_{i}^{\vee}\> s_i\right) \>\ket{v_\s}\>,
\nfr{Centre}
hence $k=\sum_{i=0}^r k_{i}^{\vee}\> s_i \in{\Bbb Z}\geq0$. On $L(\s)$ 
there is a notion of orthogonality by means of a (unique) positive
definite Hermitian form $H$ such that $H(\ket{v_\s}, \ket{v_\s}) \equiv
\langle v_\s \ket{v_\s}= 1$. Finally, $L(\s)$ can be ``integrated'' to a
representation of the Kac-Moody group $\GG$, which is then generated just
by the exponentials of the generators of
$\ggg$~[\Ref{PK},\Ref{KBOOK},\Ref{KP}].

We will use the notation $\ket{v_i}$ for the highest-weigh vector of
the ``fundamental'' representation $L(i)$ where $s_j =
\delta_{j,i}$, and $d_i$ for the corresponding derivation. In terms of
these fundamental highest-weight vectors, $\ket{v_\s}$ can be decomposed
as
$$
\ket{v_\s} \>=\> \bigotimes_{i=0}^{r}\> \bigl\{ \ket{v_i}^{\otimes s_i}
\bigr\}\>.
\nfr{Tensor}

It follows from its definition that the highest-weight vector of $L(\s)$
is annihilated by $\gg_{>0}(\s)$, and that it is an eigenvector of
$\gg_{0}(\s)$ with eigenvalues
$$
\eqalign{
& h_i\> \ket{v_\s}\> =\> s_i \> \ket{v_\s}\>, \qquad d_{\s}\> \ket{v_\s}
\>=\> 0\>, \cr
& e_{j}^-\> \ket{v_\s} \>=\> 0\quad {\rm when}\quad s_j=0\>.\cr}
\nfr{Eigen}
Then, the representation of the subgroups $\GG_+(\s)$, $\GG_-(\s)$, and
$\GG_0(\s)$ on $L(\s)$ are actually generated by exponentiating the
generators of $\gg_{>0} (\s)$, $\gg_{<0}(\s)$, and $\gg_0(\s)$,
respectively.

For simply laced affine Kac-Moody algebras, all the fundamental
integrable representations of level-one are isomorphic to the basic
representation $L(0)$, which can be realized in terms of vertex operators acting
on Fock spaces~[\Ref{KP},\Ref{VO}]. Then, the other fundamental integrable
representations of level~$>1$ can be realized as submodules in the tensor
product of several fundamental level-one representations. Moreover, the
fundamental integrable representations of non-simply laced Kac-Moody algebras
can be constructed from those of the simply laced algebras by folding
them~[\Ref{TUROLB},\Ref{MARCO}].

\references

\beginref
\Rref{FMDOS}{L.~Feh\'er, J.~Harnad and I.~Marshall, Commun. Math. Phys.
{\bf 154} (1993) 181;\newline
L.~Feh\'er and I.~Marshall, {\sl Extensions of the matrix Gelfand--Dickey
hierarchy from generalized Drinfeld--Sokolov reduction\/}, SWAT/95/61,
hep-th/9503217; \newline
H.~Aratyn, L.A.~Ferreira, J.F.~Gomes and A.H.~Zimerman, {\sl
Constrained KP Models as Integrable Matrix Hierarchies\/}, IFT-P/041/95,
UICHEP-TH/95-9, hep-th/9509096.}
\Rref{VO}{I.B.~Frenkel and V.G.~Kac, Invent. Math. {\bf 62} (1980) 28;\newline
J.~Lepowsky and R.L.~Wilson, Commun. Math. Phys. {\bf 62} (1978) 43; \newline
V.G.~Kac, D.A.~Kazhdan, J.~Lepowsky and R.L.~Wilson, Adv. in Math. {\bf 42}
(1981) 83;\newline
J.~Lepowsky, Proc. Natl. Acad. Sci. USA {\bf 82} (1985) 8295.}
\Rref{MAX}{M.R.~Niedermaier, Commun. Math. Phys. {\bf 160} (1994) 391.}
\Rref{OLTUR2} {D.~Olive, N.~Turok, Nucl. Phys. {\bf B265} [FS15] (1986) 
 469-484.}
\Rref{mm}{N.~McKay and W.A.~McGhee, Int. Jour. Mod. Phys. {\bf A8} (1993) 
 2791-2807, Erratum ibid. {\bf A8} (1993) 3830, hep-th/9208057.}
\Rref{PARK}{T.J.~Hollowood, J.~Luis Miramontes and Q-Han Park, 
Nucl.Phys. {\bf B445} (1995) 451-468, hep-th/9412062; \newline
C.R.~Fern\'andez-Pousa, M.V.~Gallas, T.J.~Hollowood and J.L.~Miramontes, {\sl
The Symmetric Space and Homogeneous sine-Gordon Theories\/}, hep-th/9606032.}
\Rref{SAVGERV}{L.A.~Ferreira, J-L.~Gervais, J.~S\'anchez Guill\'en and 
M.V.~Saveliev,  {\sl Affine Toda Systems Coupled to Matter Fields}, to appear 
in Nucl. Phys. {\bf B}, hep-th/9512105.}
\Rref{ZS}{V.E.~Zakharov and A.B.~Shabat, Funct. Anal. Appl. {\bf 13} (1979)
166.}
\Rref{HKDV}{R.~Hirota, Phys. Rev. Lett. {\bf 27} (1971) 1192.}
\Rref{LUIZ}{L.A.~Ferreira, J.L.~Miramontes and J.~S\'anchez Guill\'en,
Nucl. Phys. {\bf B449} (1995) 631-679, hep-th/9412127.} 
\Rref{DIC}{L.A.~Dickey, {\sl On tau-functions of Zakharov-Shabat
and other matrix hierarchies of integrable equations\/},
hep-th/9502063;\newline
H.~Aratyn, {\sl Integrable Lax Hierarchies, their Symmetry Reductions and 
Multimatrix Models}, 
lectures at VIII~J.A.~Swieca Summer School (Rio de Janeiro, Brazil, 1995), 
hep-th/9503211.}
\Rref{TAUTIM}{T.J.~Hollowood and J.L.~Miramontes, Commun. Math.
Phys.{\bf 157} (1993) 99.}
\Rref{GEN}{M.F.~de Groot, T.J.~Hollowood, and J.L. Miramontes, Commun. 
Math. Phys.{\bf 145} (1992) 57;\newline
N.J.~Burroughs, M.F.~de Groot, T.J.~Hollowood, and J.L. Miramontes, 
Commun. Math. Phys.{\bf 153} (1993) 187; Phys. Lett. {\bf B~277}
(1992) 89.}
\Rref{KP}{V.G.~Kac and D.H.~Peterson, {\sl 112 constructions of
the basic representation of the loop group of $E_8$\/}, in ``Symposium
on Anomalies, Geometry and Topology'' (W.A.~Bardeen and
A.R.~White, eds.), World Scientific (1985) 276.}
\Rref{DS}{V.G.~Drinfel'd and V.V.~Sokolov, J.~Sov. Math. {\bf 30}
(1985) 1975.}
\Rref{KW}{V.G.~Kac and M.~Wakimoto, {\sl Exceptional hierarchies
of soliton equations\/}, in ``Proceedings of Symposia in Pure
Mathematics'', Vol. {\bf 49} (1989) 191.}
\Rref{TAUJAP}{M.~Jimbo and T.~Miwa, Publ. RIMS, Kyoto Univ. {\bf
19} (1983) 943;\newline
E.~Date, M.~Jimbo, M.~Kashiwara, and T.~Miwa, Publ. RIMS, Kyoto Univ. 
{\bf 18} (1982) 1077.}
\Rref{HIRL}{H.~Aratyn, C.P.~Constantinidis, L.A.~Ferreira, J.F.~Gomes,
and A.H.~Zimerman, Nucl. Phys. {\bf B~406} (1993) 727-770, 
hep-th/9212086.}
\Rref{HIR}{R.~Hirota, {\sl Direct methods in soliton theory\/}, in
``Soliton'' (R.K.~Bullough and P.S.~Caudrey, eds.), (1980) 157.}
\Rref{HSG}{R.~Hirota, J.~Phys. Soc. Japan {\bf 33} (1972) 1459.}
\Rref{TIMSOL}{T.J.~Hollowood, Nucl. Phys. {\bf B~384} (1992) 523.}
\Rref{CAT}{O.~Babelon and L.~Bonora, Phys. Lett. {\bf 244~B}
(1990) 220;\newline H.~Aratyn, L.A.~Ferreira, J.F.~Gomes and
A.H.~Zimerman, Phys. Lett. {\bf B~254} (1991) 372.} 
\Rref{PK}{D.H.~Peterson and V.G.~Kac, Proc. Natl. Acad. Sci.
USA {\bf 80} (1983) 1778;\newline V.G.~Kac, {\sl
Constructing groups associated to infinite-dimensional Lie
algebras\/}, in ``Infinite dimensional groups with
applications'' (V.G.~Kac, ed.), Berkeley MSRI pub., Springer
Verlag (1985) 167.} 
\Rref{KBOOK}{V.G.~Kac, {\sl Infinite
dimensional Lie algebras\/} ($3^{rd}$ ed.), Cambridge
University Press (1990);\newline
J.~Fuchs, {\sl Affine Lie Algebras and Quantum Groups\/}, Cambridge
University Press (1992).} 
\Rref{WILa}{G.~Wilson, Phil. Trans. R.~Soc. Lond. {\bf A~315} (1985)
383; {\sl Habillage et fonctions $\tau$\/} C.~R. Acad. Sc. Paris {\bf 299 (I)}
(1984) 587.}
\Rref{WILb}{G.~Wilson, {\sl The $\tau$-Functions of the gAKNS
Equations}, in ``Verdier memorial conference on integrable systems''
(O.~Babelon, P.~Cartier, and  Y.~Kosmann-Schwarzbach, eds.), Birkhauser
(1993) 131-145.}
\Rref{BK}{M.J.~Bergvelt and A.P.E.~ten Kroode, J.~Math. Phys. {\bf 29} (1988)
1308.}
\Rref{BABELON}{O.~Babelon and D.~Bernard, Phys. Lett. {\bf B260} 81; Commun.
Math. Phys. {\bf 149} (1992) 279-306, hep-th/9111036.}
\Rref{BABT}{O.~Babelon and D.~Bernard, Int.~J. Mod. Phys. {\bf A8} (1993) 
507-543, hep-th/9206002.}
\Rref{OLTUR}{D.~Olive and N.~Turok, Nucl. Phys. {\bf B257} (1985) 277}
\Rref{UNDER}{J.W.R.~Underwood, {\sl Aspects of Non Abelian Toda Theories},
hep-th/9304156.}
\Rref{SOLSPEC}{D.~Olive, M.V.~Saveliev and J.W.R. Underwood, Phys. Lett. 
{\bf B311} (1993) 117-122, hep-th/9212123.}
\Rref{LS}{A.N.~Leznov and M.V.~Saveliev, {\sl Group Theoretical Methods for 
Integration of Non-Linear Dynamical Systems}, Progress in Physics Series, 
v. 15, Birkha\"user-Verlag, Basel, 1992.}
\Rref{TUROLA}{D.~Olive, N.~Turok and J.W.R.~Underwood, Nucl. Phys. {\bf B401} 
(1993) 663-697.}
\Rref{TUROLB}{D.~Olive, N.~Turok and J.W.R.~Underwood, Nucl. Phys. {\bf B409}
(1993) 509-546, hep-th/9305160.}
\Rref{DOV}{D.~Olive, private communication.}
\Rref{MARCO}{A.~Fring, P.R.~Johnson, M.A.C.~Kneipp, and D.I.~Olive, Nucl. Phys.
{\bf B430} [FS] (1994) 597-614; \newline
M.A.C.~Kneipp, and D.I.~Olive,  Commun. Math. Phys. {\bf 177} (1996) 561-582.}

\endref

\ciao